\documentclass[a4paper,fleqn,usenatbib]{mnras}

\usepackage[T1]{fontenc}
\usepackage{ae,aecompl}

\usepackage{graphicx}	
\usepackage{amsmath}	
\usepackage{amssymb}	
\usepackage{morefloats}

\newcommand{\htwoo}{H$_2$O}

\newcommand{\de}{$^{\circ}$}
\newcommand{\methane}{CH$_4$}
\newcommand{\ms}{m~s$^{-1}$}

\title[GCM predictions for New Horizons temperatures]{An atmospheric general circulation model for Pluto with predictions for New Horizons temperature profiles}

\author[A. M. Zalucha]{
Angela M. Zalucha,$^{1}$\thanks{E-mail: azalucha@seti.org}
\\
$^{1}$SETI Institute, Mountain View, California, 94043, USA
}

\date{Accepted XXX. Received YYY; in original form ZZZ}

\pubyear{2015}

\begin{document}
\label{firstpage}
\pagerange{\pageref{firstpage}--\pageref{lastpage}}
\maketitle

\begin{abstract}
\textbf{Results are presented from a 3-D Pluto general circulation model (GCM) that includes conductive heating and cooling, non-local thermodynamic equilibrium (non-LTE) heating by methane at 2.3 and 3.3 microns, non-LTE cooling by cooling by methane at 7.6 microns, and LTE CO rotational line cooling.  The GCM also includes a treatment of the subsurface temperature and surface-atmosphere mass exchange.  An initially 1 m thick layer of surface nitrogen frost was assumed} such that it was large enough to act as a large heat sink (compared with the solar heating term) but small enough that the water ice subsurface properties were also significant. Structure was found in all three directions of the 3-D wind field (with a maximum magnitude of order 10 m/s in the horizontal directions and 10$^-5$ microbar/s in the vertical direction).  Prograde jets were found at several altitudes.  The direction of flow over the poles was found to very with altitude.  Broad regions of up-welling and down-welling were also found.  Predictions of vertical temperature profiles are provided for the Alice and REX instruments on New Horizons, while predictions of light curves are provided for ground-based stellar occultation observations.  With this model methane concentrations of 0.2\% and 1.0\% and 8 and 24 microbar surface pressures are distinguishable.  For ground-based stellar occultations, a detectable difference exists between light curves with the different methane concentrations, but not for different initial global mean surface pressures.
\end{abstract}




\section{Introduction}
The modelling of Pluto's atmosphere began with its discovery in 1988 by the stellar occultation technique~\citep{elliot:1989}.  These ranged from mathematically convenient models, such as that of \citet{elliot:1992}, which assumed that the temperature dependence with height was a power law, to those of \citet{yelle:1989}, which used conduction and radiative transfer in the non-local thermodynamic equilibrium (non-LTE) regime to describe heating by \methane~at 3.3~$\mu$m and cooling by \methane~at 7.6~$\mu$m in order to predict temperature with height.  This type of radiative-conductive model was expanded to an additional \methane~line at 2.3~$\mu$m and LTE cooling by CO rotational line emission by~\citet{strobel:1996}, and a troposphere with a user-specified height was added by~\citet{zalucha:2011b}.  \citet{zhu:2014} further expanded this model to the upper atmosphere of Pluto by adding the processes of atmospheric escape, UV heating, and adiabatic cooling due to atmospheric expansion.  Meanwhile, a technique was developed by~\citet{elliot:2003b} to invert stellar occultation light curve flux vs. time to temperature vs. height.  This algorithm required a boundary condition at the ``top'' of the atmosphere (usually $\sim$~100~km altitude, depending on the surface radius assumed), which was then used to integrate downward in the atmosphere (closer to the occultation midtime) assuming hydrostatic equilibrium, the ideal gas law, the \textbf{linear} proportionality of number density to refractivity, and the equations of light in the geometric limit of optics. 

The general result of 1-D modelling efforts~\citep[e.g.,][]{yelle:1989,hubbard:1990,elliot:1992,strobel:1996,hansen:1996,elliot:2003b,zalucha:2011a,zalucha:2011b,young:2012,young:2013,zhu:2014}, spectroscopy~\citep[e.g.,][]{owen:1993,tryka:1993,lellouch:2009,greaves:2011,lellouch:2011,holler:2014}, and stellar occultation observations~\citep[e.g.,][]{elliot:1989,elliot:2003a,sicardy:2003,pasachoff:2005,elliot:2007,eyoung:2008,person:2008} was that Pluto's atmosphere was primarily composed of N$_2$, \methane, CO, and ethane.  Pluto's surface temperature was $40\pm2$~\citep{tryka:1994}.  Before the flyby of Pluto by NASA's New Horizon's spacecraft on 2015 July 14, evidence was inconclusive about the presence of a troposphere and its depth~\citep{stansberry:1994,zalucha:2011b}.  Preliminary results from New Horizons show a shallow troposphere confined to the boundary layer~\citep{stern:2015}.  Otherwise, the lower atmosphere (referred to in Pluto literature as the stratosphere) is characterized by a sharp temperature inversion (temperature increasing with height) of $\sim$~5~K~km$^{-1}$).  The temperature becomes mostly isothermal at altitudes of $\sim$ 100--600~km.  Preliminary results from New Horizons have determined the surface radius of Pluto to be 1187$\pm$4~km and the global mean surface pressure to be 10~$\mu$bar~\citep{stern:2015}\footnote{1 $\mu$bar = 0.1 Pa}.  Prior to the flyby, analysis of stellar occultation data and \methane~spectral lines estimated the global mean surface pressure to be between 6 and 24~$\mu$bar and changing with time \citep{lellouch:2009,zalucha:2011a,hansen:1996,young:2013}.  Specifically, the pressure underwent an increase, perhaps by as much as a factor of two from 1988 to 2003, and remained mostly constant after 2003.  Based on modelling studies, it was hypothesized that Pluto's atmosphere sublimates and condenses in vapour pressure equilibrium with surface ice as Pluto orbits the Sun.  Preliminary results from New Horizons also show possible evidence for sublimation~\citep{stern:2015}.

An important aspect of those 1-D models is that they ignored the transport of heat by wind and adiabatic heating and cooling due to downward and upward motions of air.  There is no \textit{a priori} reason to suspect that Pluto's atmosphere lacks wind~\citep[and preliminary results from New Horizons have tentatively identified aeolian surface features][]{stern:2015}; thus, modellers have begun investigating wind using general circulation models (GCMs) for both Pluto and Triton~\citep{mueller-wodarg:2001,vangvichith:2011,michaels:2011,miller:2011,zalucha:2012,zalucha:2013,toigo:2015}.  GCMs solve the Navier-Stokes equations, continuity of mass and energy, and an equation of state on a global scale with a horizontal resolution of a few degrees and a vertical range of several scale heights.  They are advantageous because they solve the system of equations describing geophysical fluid dynamics, but disadvantageous because they are computationally expensive.  

The list of citations above represent six different efforts at transforming terrestrial GCMs to Pluto and/or Triton (which have similar atmospheres) using knowledge of other solar system bodies with atmospheres where applicable.  The mystery of Pluto's wind structure at present cannot be answered observationally, since the effects of wind can only be ascertained remotely \textbf{in} certain situations (e.g. cloud movements, ocean roughness, and/or surface dunes).  Thus, GCMs are extremely important for \textbf{investigating} Pluto's atmospheric circulation with full global and temporal coverage, which in turn affects its 3-D temperature field, surface pressure, and transport of volatiles.  Currently there is no common consensus among these GCMs as to Pluto's circulation.  Even with the same input parameters, GCMs may produce different results due to differing numerical schemes at the fundamental level.  In more mature fields such as Earth or Martian GCMs, intercomparison studies are conducted that strip GCMs down to their basic components.  The field of Pluto GCM research is simply too new to have had any of these studies performed (though, it appears there is good reason to do so in the future).

The two main instruments on New Horizons that observed Pluto's lower atmosphere are Alice~\citep{stern:2008} and the Radio science Experiment~\citep[REX,][]{tyler:2008}.  Alice is an ultraviolet imaging spectrometer.  Alice has two modes: one that observed the airglow emission (a collection of processes that occur in the upper atmosphere at night) of Pluto's atmosphere and one that performed occultation experiments.  The particular configuration of the Pluto GCM (PGCM) used in this study cannot predict airglow; however, the latter capability is of interest here.  The first type of occultation occurred when Pluto occulted the Sun, such that sunlight shined through Pluto's atmosphere and was received by New Horizons.  The second type of occultation was similar, except Pluto occulted a background star.  These types of occultations, described in Section~\ref{se:lc}, allow temperature to be derived from light intensity based on the attenuation of light due to differential refraction of light through an atmosphere.  Although the flyby has already occurred, it will take some time for the data to be transmitted to Earth and subsequently reduced to a form that can be compared with physical models.  The predictions of this paper are therefore relevant.

REX performed a different type of occultation experiment.  REX received radio waves sent from Earth as they passed through Pluto's atmosphere.  The observed frequency modulation, again due to differential refraction of light through an atmosphere, also allowed for the retrieval of temperature with altitude.  A key point of REX's observations was \textbf{that} the temperature was measured down to the surface (and the surface radius of Pluto itself was for the first time well-constrained), something that ground-based stellar occultations cannot do.  For Pluto, ground-based stellar occultations measure down to about 20 km altitude~\citep[e.g.][]{zalucha:2011a}.  This inability to measure temperature to the surface has to do with the fact that the S/N ratio near the midtime of the light curve becomes too low, and these observations essentially ``run out of light.''  The fundamental difference between UV, solar, or IR occultations (as typically performed in ground-based studies) vs. radio occultations is the latter provided a temperature measurement near the surface for the first time.

In this paper, the development of the \citet{zalucha:2012} (hereinafter known as ZG12) and \citet{zalucha:2013} (hereinafter known as ZM13) line of Pluto GCMs (PGCMs) is continued.  The former was a 2-D PGCM (latitude, height, and time) and the latter was a 3-D PGCM (longitude, latitude, height, and time).  The main difference with adding the longitude direction was that (transverse) waves in the longitudinal direction were allowed.  This extra dimension allowed for the formation of thermal tides, Rossby waves, Kelvin waves, and gravity (buoyancy) waves.  These waves are physically real and can transport momentum and heat in all three dimensions as well as dissipate artificial noise produced by the model.  ZM13 predicted a prograde jet at the equator with a maximum magnitude of 10--12~\ms.  The surface pressure predicted for 1988, when compared with observed stellar occultation light curves, was 8$\pm$4~$\mu$bar.  Likewise for 2002 and 2006 it was 20$\pm$4~$\mu$bar, and for 2007 it was 16$\pm$4~$\mu$bar.  All years had a greater than 1\% concentration of \methane needed to fit the data, which was higher than results derived from modelling of spectroscopic observations~\citep[0.5\%][]{lellouch:2009} and the preliminary value from New Horizons of 0.25\%~\citep{stern:2015}.

\textbf{The \textbf{primary} upgrade to the PGCM of ZM13 was to replace the radiative-conductive model with that of  \citet{strobel 1996}, which includes a more sophisticated treatment of the \methane~3.3 and 7.6~$\mu$m lines, the addition of the \methane~2.3~$\mu$m line, and the effects of CO cooling.  Furthermore, a new method of equilibrating the atmosphere during the initial spin-up time~\citep[developed independently of][]{toigo:2015} has been implemented.  A surface-atmosphere mass exchange prescription and subsurface and surface model have also been introduced, but have not been extensively tested (a subject of future work).  }In Section~\ref{se:model} the model is described, in Section~\ref{se:setup} the simulation setup is described, in Section~\ref{se:results} simulation results are shown for key variables (temperature and wind) and interpretation of the model results is provided, in Section~\ref{se:lc} predictions of visible-wavelength stellar occultation light curves that ground-based observers would have expected to see at the time of the New Horizons encounter are made, and in Section~\ref{se:nh} predictions of temperature profiles that were be observed by Alice and REX (but not yet downloaded to Earth) are presented.  The date for the ground-based predictions is arbitrary, as at present no ground-based stellar occultations were known occur at that time.  At present, it is not known if Pluto's atmosphere will stay in this state for several decades (given Pluto's 248 Earth-year long year and sluggish atmosphere) or if there will be a sudden atmospheric collapse \citep{young:2013,olkin:2015}.  In Section~\ref{se:conclusions}, a discussion is provided about speculations future regarding multi-year simulations.

\section{The Model}\label{se:model}
The PGCM is based off of the Massachusetts Institute of Technology (MIT) atmospheric GCM~\citep{marshall:1997}.  The model solves the primitive equations of geophysical fluid dynamics in 3-D using the finite volume method on an Arakawa C-grid.  These equations are: horizontal momentum
\begin{eqnarray}
\frac{Du}{Dt}+\frac{\partial \Phi}{\partial x} - fv&=&F_x \label{eq:xmom}\\
\frac{Dv}{Dt}+\frac{\partial \Phi}{\partial y} +fu &=& F_y,\label{eq:ymom}
\end{eqnarray}
where $D[\;]/Dt=\partial[\;]/\partial t+\nabla \cdot [\;]$ is the total derivative, $u$ is the velocity (relative to the body's surface) in the longitudinal direction, $v$ is the velocity in the latitudinal direction, $x$ is linear distance in the eastward direction, $y$ is linear distance in the northward direction, $t$ is time, $\Phi$ is the geopotential ($gz$, where $g$ is the gravitational acceleration at the surface and $z$ is height), $f=2\Omega \sin\phi$ is the Coriolis parameter (where $\Omega$ is the body's rotation rate and $\phi$ is latitude), and $F_x$ and $F_y$ are external forcings, in this case friction and velocity drag; vertical momentum in the hydrostatic approximation
\begin{equation}
\frac{\partial\Phi}{\partial p} = -a,
\end{equation}
where $p$ is pressure and $a$ is the inverse of density; an equation of state, in this case the ideal gas law
\begin{equation}
pa= R T,
\end{equation}
where $R$ is the specific gas constant (the universal gas constant divided by the weighted average of the molecular weights of the atmospheric constituents) and $T$ is temperature; mass equation
\begin{equation}\label{eq:mass}
\frac{\partial u}{\partial x}+\frac{\partial v}{\partial y}+\frac{\partial \omega}{\partial p}=M,
\end{equation}
where $\omega$ is the vertical velocity in pressure coordinates and $M$ is the mass source or sink term due to condensation or sublimation; and an energy equation
\begin{equation}
c_v\frac{DT}{Dt}+p\frac{Da}{Dt}=J\label{eq:energy},
\end{equation}
where $c_v$ the specific heat at constant volume and $J$ is an external heating and cooling term described below.

\begin{figure*}
 \noindent\includegraphics[width=1.0\textwidth]{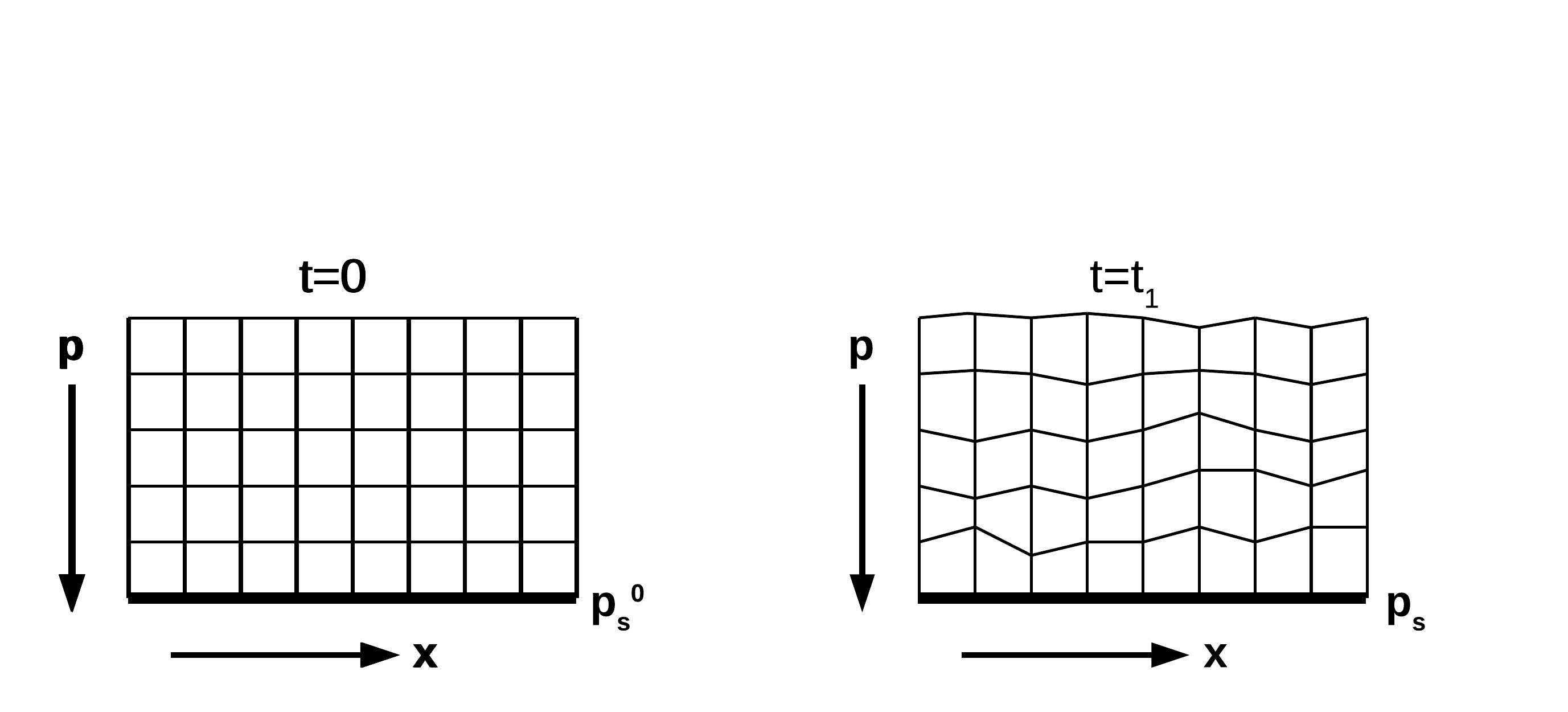}
 \caption{\label{fg:eta_coords} Schematic of $p^*$ coordinate system.  For the sake of brevity, only the $x$-axis is shown but the same process applies to the horizontal $y$-axis.  At time $t=0$ (left panel), pressure values are defined on the vertical grid such that they are the same globally for a given level.  The initial surface pressure ($p_s^0$) is also globally constant.  As the model evolves to some time $t_1$ later (right panel), air has moved between the boxes (due to the physical and thermodynamic forces acting on them) and their vertical boundaries have changed.  The surface pressure $p_s$ is the sum of the height of the column, as depicted in the figure.  The variations in the vertical boundaries are usually small. }
   \end{figure*}

In the MIT GCM, the vertical coordinate used is not exactly pressure as described in equations~(\ref{eq:xmom})--(\ref{eq:energy}), but instead a modified pressure coordinate referred to as $p^*$~\citep{adcroft:2004}, or more commonly known as $\eta$ in atmospheric literature.  The definition of this coordinate is
\begin{equation}
p^*(x,y,t)=p(x,y,t)\frac{p^0_s(x,y)}{p_s(x,y,t).}
\end{equation}
For a given $x$, $y$, and $p$ grid box, the vertical boundaries of the box may move up or down, thus shrinking or expanding the box.  Figure~\ref{fg:eta_coords} shows the grid at the start of integration.  Initially, the vertical boundaries of each vertical grid point lie on the same pressure level (left panel).  The initial surface pressure $p^0_s$ is globally constant.  As the model evolves (right panel), the boxes vertically shrink or expand depending on the atmospheric dynamics.  Note that the surface pressure $p_s$ changes with time and in the figure is represented by the column height.  In practice, the fluctuations of the vertical boundaries are small ($\le \left|0.1\right|$) compared with their original position.  Table~\ref{tb:vgrid} shows the values of the initial pressure levels (the \textbf{case} numbers will be defined in Section~\ref{se:setup}).  The MIT GCM includes two sets of staggered grids in vertical, the $p_f$ grid (``f'' for faces) and  $p_c$ grid (``c'' for centres).  The $p_f$ grid has 31 points, with the top point being at zero pressure and the bottom point being the surface pressure.  The $p_c$ grid points (of which there are 30) are located halfway between the $p_f$ grid points.  The number of vertical grid points was chosen such that vertical structure could be sufficiently resolved while making the model computationally feasible.  The reason for the two grids is that for the purposes of taking derivatives; $T$, $u$, $v$, and $\Phi$ are located on the $p_c$ grid and $\omega$ is located on the $p_f$ grid.  This definition leads us to the model's upper and lower boundary conditions: $\omega=0$ (air cannot flow out of the atmosphere or into the ground).  As is common in GCMs, more points are located near the surface since the dynamics are more intricate (owing to boundary layer physics).  The model is insensitive to the precise position of the points.  This practice also makes sense for Pluto given the large vertical temperature gradient in the lower atmosphere.

   \begin{table}
\centering
\caption{Pressure values for vertical grids used in the PGCM}\label{tb:vgrid}
\begin{tabular}{cll}
\hline

Level    & \textbf{Cases}  1 and 3              & \textbf{Cases}  2 and 4              \\
         &  ($\mu$bar)  & ($\mu$bar)  \\
\hline
1 & 8.000000 & 24.000000 \\
2 & 7.860086 & 23.580258 \\
3 & 7.651933 & 22.955798 \\
4 & 7.351503 & 22.054509 \\
5 & 6.924851 & 21.414532 \\
6 & 6.333587 & 19.887658 \\
7 & 5.543393 & 17.815471 \\
8 & 4.583648 & 15.190562 \\
9 & 3.693600 & 12.415871 \\
10 & 2.927441 & 9.0931562 \\
11 & 2.291569 & 7.828516 \\
12 & 1.778336 & 6.105858 \\
13 & 1.371895 & 4.715346 \\
14 & 1.054218 & 3.639138 \\
15 & 0.807902 & 2.793179 \\
16 & 0.617827 & 2.138594 \\
17 & 0.471519 & 1.634020 \\
18 & 0.358990 & 1.245765 \\
19 & 0.272382 & 0.947059 \\
20 & 0.205579 & 0.716942 \\
21 & 0.153848 & 0.539140 \\
22 & 0.113559 & 0.401111 \\
23 & 0.081964 & 0.293284 \\
24 & 0.057051 & 0.208522 \\
25 & 0.037467 & 0.141777 \\
26 & 0.024502 & 0.089876 \\
27 & 0.016861 & 0.051204 \\
28 & 0.004944 & 0.024945 \\
29 & 0.001969 & 0.009769 \\
30 & 0.000317 & 0.002829 \\
31 & 0.000000 & 0.000000 \\

\hline
\end{tabular}
\end{table}

The forcing terms in equations~(\ref{eq:xmom}) and (\ref{eq:ymom}) include two effects.  The first is surface friction, represented by a simple drag law $-k_vu$ and $-k_vv$, where
\begin{equation}
k_v=k_f\max\left(0,\frac{p-p_b}{p_s-p_b}\right).\label{eq:kv}
\end{equation}
Here $p_b$ is the pressure top of the frictional layer and $p_s$ is the surface pressure.  The variable $k_f$ is an an empirical constant and acts as a type of bulk coefficient of friction.  In more involved treatments of the boundary layer, this forcing term is replaced with yet more complex empirical treatments of the surface-air interaction, because the scale of surface features is usually much smaller than the GCM horizontal grid.  The only bodies for which we have empirical data to estimate this constant are Earth and Mars.  For Earth, this parametrization (equation~\ref{eq:kv}) was first used in the simple Earth model of \citet{held:1994} with a value of an inverse Earth day.  Similar values have been used for Mars~\citep[e.g.][]{lewis:1996,nayvelt:1997}).  Since there there is no empirical data for Pluto, nor to the author's knowledge has there been a published study on the effect of changing surface pressure on $k_f$, it is left as one inverse Earth day here.  Furthermore $p_b/p_s=0.7$ is allowed in accordance with the original Earth (and Mars) model, as we have no data to constrain it for Pluto.   Ideally, sensitivity tests should be performed on the effect of the value of $k_f$ and other similarly unknown parameters; however, due to the large computational expense of Pluto simulations, it will be left to future work.  Sensitivity tests are performed on other major quantities in Section~\ref{ss:u}.

The second forcing effect in the horizontal momentum forcing term is a "frictional" term at the model top.  In a model atmosphere, a rigid lid (i.e., $\omega$=0) is applied at the model top that does not exist in the real world.  This model top can cause spurious wave reflections off the lid that are non-physical.  In the real atmosphere, upward propagating gravity waves break \textbf{in the} upper atmosphere, depositing momentum and creating a drag on the flow.  The properties of these waves can vary with time and space.  Because we have practically no data on this process for Pluto~\citep[but see][]{person:2008,hubbard:2009}, it is difficult to predict theoretically, and waves are smaller in scale than the grid boxes, model simulations were performed with a constant drag coefficient set at different orders of magnitude and spanning a different number of top levels, a common practice in simple GCMs.  It was tested as to which configuration does not damp the upper atmosphere so much that it acts like part of the rigid lid, while preventing wave reflections.  In general, the coefficient's magnitude must increase with height.  It was found that the best value for the coefficients for drag in the top three model levels that are from the top downwards: 0.0001013799, 0.0000337933, 0.0000112644 s$^{-1}$.  In addition, an 8th-order Shapiro filter is included in the momentum forcing term to damp subgridscale, numerical noise.

Surface topography is assumed to be flat.  It has been found from New Horizons data that mountains locally rise 2 to 3~km above the surrounding orography.  It is also hypothesized that the informally named Sputnik Planum is a basin filled with CO ice bordered by mostly higher terrain.  As the height of the mountains is much less than the scale height of 50~km, they can be assumed to fall under the domain of the surface friction term in the PGCM.  It would be interesting to see how these mountains affect the flow at low-levels in a regional (i.e. mesoscale) simulation.

The energy forcing term in Eq.~\ref{eq:energy}, i.e., \textbf{atmospheric} heating and cooling terms, were taken from \citet{strobel:1996}. The 1-D heat equation is 
\begin{equation}\label{eq:dtdt}
c_p\rho(r,t)\frac{\partial T(r,t)}{\partial t} = \frac{1}{r^2}\frac{\partial}{\partial r}\left(r^2 K \frac{\partial T(r,t)}{\partial r}\right) + R_{net}(r,t),
\end{equation}
where $c_p$ is the specific heat at constant pressure, $\rho(r,t)$ is the air density, $K$ is the thermal conductivity, and $R_{net}(r,t)$ is the heating rate.  Note that the terms in Eq.~\ref{eq:dtdt} are the external heating terms, $J$, in Eq.~\ref{eq:energy}.  For Pluto a primarily N$_2$ atmosphere with smaller amounts of \methane~and CO was assumed.  For an N$_2$, \methane, and CO mixture,
\begin{equation}
K(r,t)=K_o T(r,t)^{\alpha}
\end{equation}
where $K_o=5.63\times10^{-5}$~J~m$^{-1}$~s$^{-1}$~K$^{-\left(\alpha+1\right)}$ and $\alpha=1.12$~\citep{hubbard:1990}.  The most mathematically complex part of the model is the specification of $R_{net}(r,t)$, which is the sum of the non-LTE heating rate by solar near-IR absorption in the \methane~2.3 and 3.3~$\mu$m vibrational bands, the non-LTE cooling rate due to the \methane~7.6~$\mu$m vibrational band, and the LTE cooling rate by CO rotational line emission\footnote{New Horizons found evidence of C$_2$H$_2$, which is currently being added to the $R_{net}(r,t)$ term as a non-LTE cooling line at 13.7~$\mu$m.  Moreover, the effects of \methane~non-LTE heating at 1.67~$\mu$m have also been identified as a key heating source and is also being added to the 1-D heating and cooling prescription (X. Zhu, private communication).}.  In the ZG12 specification of the model, the disk-averaged insolation was been replaced with a fully longitudinally, latitudinally, and seasonally varying insolation.  Finally, a stratosphere only configuration \citep[i.e., no troposphere is present as in][]{zalucha:2011b} was assumed, which is broadly consistent with the findings of New Horizons~\citep{stern:2015}.

The PGCM is capable of undergoing both mass and energy changes due to freezing and sublimation by the primary atmospheric constituent (as $\beta$-phase N$_2$ ice), represented by $M$ in equation~(\ref{eq:mass}).  If the atmospheric or surface temperature falls below the freezing point, the amount of heat needed to return it to the freezing point is calculated, and this heat is provided by latent heating due to mass freezing out of the atmosphere.  This mass is instantaneously removed from the atmosphere and added to the surface ice inventory.  In the case of a surface temperature below freezing, the mass is taken out of the lowest model layer and added to the surface ice inventory.  Here the heat exchange is implicit, and treated separately from the radiative-conductive heating equation~ (\ref{eq:dtdt}), following the relation
\begin{equation}
Lm=\rho A \Delta z c_p (T_{fs}-T_s)
\end{equation} 
where $L=253000$~J~kg$^{-1}$ is the latent heat of sublimation, $m$ is the amount of mass accumulated, $A$ is the horizontal area of the grid box, $\Delta z$ is the height of the lowest atmospheric layer, and $T_s$ is the surface temperature (here below the frost point at the surface, $T_{fs}$).  What the above equation states is that enough latent heat will be supplied to bring the surface temperature back up to the frost point.  If the surface is ice-covered and the temperature is predicted to be greater than the freezing point, the excess energy is used to sublimate an amount of mass such that the surface temperature returns back down to the freezing point.  The sublimated mass is added to the lowest model layer.

There is an additional subtlety to the sublimation and deposition at the surface.  As mass is subtracted from the atmosphere, the frost temperature decreases, and the amount of frost needed to accumulate on the surface to bring the surface temperature back to the frost point decreases.  Thus a feedback develops.  For a full treatment of this effect, an iterative procedure is needed.  However, it has been determined that the error in ignoring this process is insignificant.

The PGCM of ZM13 assumed that the surface temperature was fixed at the N$_2$ surface temperature, i.e. that there was an infinite reservoir that could absorb heat at the surface.  Here the surface and subsurface temperatures are allowed to vary.  The findings of \citet{stern:2015} have recently shown that Pluto's surface ice distribution (N$_2$, \methane, and CO) is very non-uniform.  At the time of the simulations in this work, it was not known that this was the case\textbf{, and} an entirely N$_2$ ice surface was assumed based on ground-based observations~\citep{owen:1993}.  It has been assumed that the thickness of the surface N$_2$ ice is a ``surface veneer'', which was also concluded by \citet{stern:2015}.  A value of 1~m was chosen for the surface ice thickness.  Appendix~\ref{se:appendixA} describes the equations of the surface and subsurface model in detail.

 In the horizontal, a cubed-sphere grid~\citep{adcroft:2004b} with 32 $\times$ 32 points per cube face is used, equivalent to a grid spacing of 2.8\de~ at the equator.  Compared to the more common cylindrical projection grid (i.e., a latitude and longitude grid), this type of horizontal grid eliminates singularities at the poles that force northward and southward winds to zero and removes the requirement for artificial Fourier filtering in the high latitudes (in order to maintain a practical time step).  Lastly, a surface radius of 1180~km is assumed~\citep{zalucha:2011a}.  The effect on the model results from using this value compared with the the recently measured New Horizons value of 1187$\pm$4~km is quite insignificant.  The surface radius comes into play in insolation terms, gravitational acceleration terms, and the calculation of stellar occultation light curves.  A difference by as much 50~km \citep[as would be the case if the surface radius were the value of the lowest value in the literature, 1137$\pm$7][]{tholen:1990,reinsch:1994} is negligible compared with Pluto's distance from the Sun (tens of AU).  Likewise, compared with Pluto's radius, an error of 50~km is small and would also not significantly impact the gravitational acceleration at the surface.  However, as shown by~\citep{zalucha:2011a}, an error of 50~km would be significant when calculating stellar occultation light curves.  The present difference of $\sim$7~km has a slight effect on the calculation of stellar occultation light curves.

\section{Model Setup}\label{se:setup}
For this work, the following Pluto pole convention has been adopted: the north pole is in the same hemisphere as the Sun's north pole and longitude increases eastward~\citep[see][for a detailed discussion of Pluto pole conventions]{zangari:2015}.  In that sense, Pluto rotates in retrograde compared to Earth and most Solar System planets.  As such, prograde flow will be westward and plotted as positive values.  Other directions remain unchanged.  Note this convention is the opposite of both the New Horizons team convention and the current IAU convention.  From the years 1985 to 2015, the subsolar latitude (or in other words Pluto's ecliptic longitude) changes in accordance with Pluto's orbit for a given date.  Thus, seasonal changes in insolation are accounted for in the model.  On this date at the time of the New Horizons flyby, the subsolar longitude was 32\de~and the subsolar latitude was 38\de~. 

The atmosphere of Pluto is thermally ``sluggish" in that the radiative-conductive time constant is long, or in other words responds slowly to changes in radiative-conductive forcing~\citep{strobel:1996,eyoung:2008}.  The time-scale for this forcing $\tau_{rc}$ may be calculated as \citep[e.g.,][]{showman:2008}
\begin{equation}
\tau_{rc}=\delta T\frac{c_p\rho}{dF/dz},
\end{equation}
where $F$ is the net vertical radiative-conductive flux and $\delta T$ represents a Gaussian perturbation to a radiative-conductive temperature profile at some altitude.  While the choice of $\delta T$ is somewhat arbitrary, $dF/dz$ scales accordingly, and so by examining a wide range of perturbation amplitudes and thicknesses, it is usually possible to gain at least an order of magnitude estimate, if not better (fractions of Earth days).  The time-scale for Pluto in the lower atmosphere is of order of a few weeks, while the time-scale in the upper atmosphere can be a significant fraction of an Earth year.  

For Mars and Earth GCMs, where the radiative time constant is two to four Earth days and tens of days, respectively, conduction is ignored, and the initial temperature structure may be set to any value as long as it is not too pathological, and the model will respond and equilibrate over a reasonable amount of wall clock time.  For \textbf{Pluto's} atmosphere, simulations starting with global isothermal temperatures of 80 K and globally quiescent winds show atmospheric temperature changes of no more than 1 K over 40 Earth years of integration time.  Since a nearly constant temperature atmosphere is far from the radiative-conductive equilibrium solution and observations, it suggests that at least that these initial conditions are far from the spun-up state.  

To initialize this three-dimensional model from this isothermal state would be prohibitively expensive from a computational standpoint.  The radiative-conductive equilibrium (i.e. steady state, quiescent wind) is then a logical choice for an initial condition as it may be quickly calculated and is assumed to be near the fully spun-up, quasi-equilibrium state.  However the temperature gradients in this configuration are so strong that a prohibitively small time step is required in the initial stages for stability. (GCMs in typical usage do not employ variable time stepping methods, since this practice can result in different amounts of solar energy being received from one day to the next.)  A hybrid method to spin-up the model has been developed as follows.  Note that this method is conceptually the same as that of \citet{toigo:2015}; however, their paper did not appear in press until long after the simulations in this paper were complete, so the method was independently arrived at by both groups.

First, the initial temperature was set to a constant value globally with quiescent winds.  The surface and subsurface temperatures were held constant and the \textbf{surface-atmosphere mass exchange} scheme was disabled.  Then, the technique of Newtonian relaxation was used to thermally force the model in place of the direct heating and cooling rates from the \citet{zhu:2014} model.  Newtonian relaxation is a method by which the forcing term in the energy equation goes as
\begin{equation}
\frac{\partial T}{\partial t}=-\frac{\left (T-T_{eq}\right)}{\tau},
\end{equation}
where $T_{eq}$ is the radiative-conductive equilibrium temperature and $\tau$ is the radiative-conductive time constant.  A value of $\tau=30$ Earth days has been found to require a weak enough forcing while affording a minimal amount of wall clock time.  The model was run in this configuration for 90 model Earth days.  Then, using the MIT GCM's restart feature, the model was restarted, now with the heating and cooling rates specified directly from \citet{zhu:2014} (i.e., not Newtonian relaxation), with the surface and subsurface evolution, and the \textbf{surface-atmosphere mass exchange} scheme.  From then the model was run from the year 1985 to 2015, i.e., spanning the observational epoch of stellar occultation observations and NASA's New Horizons Pluto flyby.

As also recognized by \citet{toigo:2015}, current computing power is not enough to equilibrate a PGCM with both an atmospheric and surface component.  2-D surface models, such as that of \citet{hansen:1996} have shown that just the subsurface/surface alone (with a 1-layer atmosphere) takes up to four Pluto years (about one thousand Earth years) of simulation time to equilibrate.  ZM13 showed that vertical velocity in the atmosphere was very weak, most likely due to the large increase in temperature with height in the stratosphere, which makes the atmosphere extremely stable against vertical motions.  It was thus reasoned that any flow that arose from pressure differences due to sublimating or condensing surface ice would be confined to the lowest atmospheric layer and would thus be decoupled from the atmosphere aloft.  Without PGCM simulations results that started with an arbitrary N$_2$ frost distribution and spanned many Pluto years, it cannot be known whether or not the surface ice distribution is accurate, and thus it will not be focused on for the remainder of this paper.  \citet{toigo:2015} on the other hand used a different approach; they ran a 2-D surface model, similar to as that of \citet{hansen:1996}, for many Pluto years, and used that as their initial surface condition.  Again, without PGCM simulations initialized alongside a surface model spanning many years, it cannot be known if this method gives accurate results either.  Similarly, the computational expense limits extensive sensitivity tests of the not well-constrained parameters, as is the case in many planetary GCMs.  Such tests will have to wait until a future paper. 

In this study, four different plausible situations are considered based on analysis of stellar occultation light curves~\citep[e.g.,][]{zalucha:2011a}, long-term surface models~\citep[e.g.,][]{hansen:1996,young:2013}, spectroscopic data~\citep[e.g.,][]{lellouch:2009}, and direct imaging~\citep[e.g.,][]{buie:2010}.  
Table~\ref{tb:parameters} shows the parameters of each experiment (each simulation will be referred to by its simulation number hereinafter).  At the time the simulations were carried out, the values of global mean surface pressure and \methane~concentration spanned values predicted from ground-based results.  Case 3 will be the primary topic of discussion for the remainder of this paper, because it is the closest the values observed by New Horizons~\citep{stern:2015}.  The results of the other cases are still relevant, as they may represent the state of Pluto's atmosphere in the recent past (or perhaps near future), but can be found in Appendix~\ref{se:appendixB}.

\begin{table}
\centering
\caption{Surface parameters of PGCM simulations}\label{tb:parameters}
\begin{tabular}{ccc}
\hline

\textbf{Case}   & \methane      & Surface              \\
number & concentration (\%) & pressure ($\mu$bar)  \\
\hline
1& 1& 8  \\
2& 1& 24 \\
3 & 0.2& 8 \\
4 & 0.2& 24 \\

\hline
\end{tabular}
\end{table}

\section{PGCM Results and Discussion}\label{se:results}
\subsection{Temperature}\label{ss:pgcm_t}
\begin{figure*}

 \noindent\includegraphics[width=1.0\textwidth]{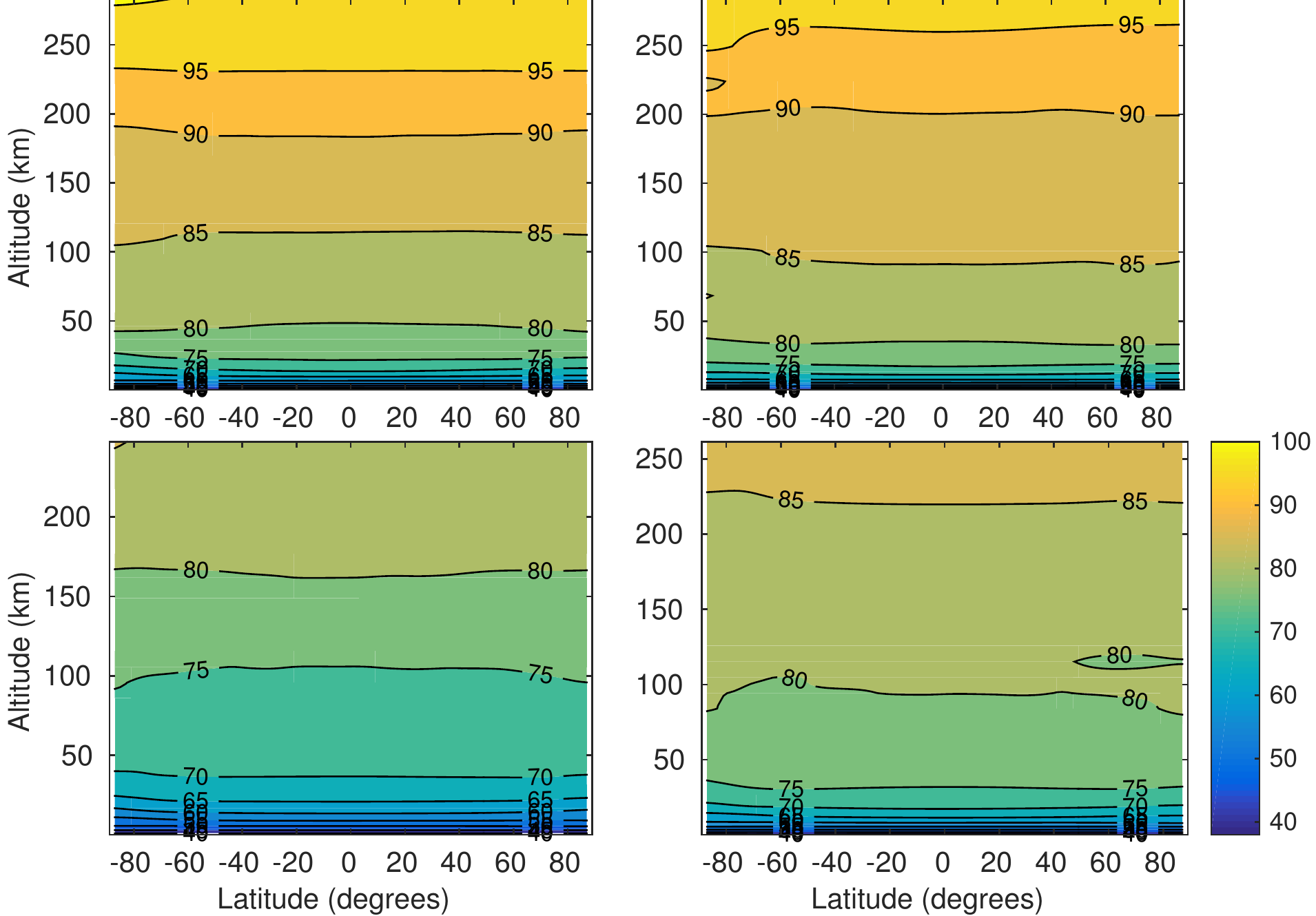}

 \caption{\label{fg:t1}  Longitudinally averaged temperature (K) for 2015 July 14.  Top left:  Case 1, top right: Case 2, bottom left: Case 3, bottom right: Case 4. The horizontal temperature gradient is weak compared with other Solar System bodies with atmospheres.  Cases 1 and 2 are the warmest temperature due to the higher \methane~concentration (which has a net heating effect on the atmosphere) and the fact that altitude is plotted on the ordinate rather than pressure.  Likewise, for Cases 3 and 4, the temperature is colder for a given altitude than in Cases 1 and 2 because of the lower \methane~concentration.}

  \end{figure*}

The atmospheric temperature shows little longitudinal (i.e. diurnal) variation.  This property is to be expected since \citet{eyoung:2008} and ZM13 found that the time-scale for radiative-conductive forcing is much longer than a Pluto day.  In Fig.~\ref{fg:t1} the longitudinally averaged temperature as a function of altitude and latitude are shown for 2015 July 14.  For each longitudinally averaged panel, the altitude values are calculated by assuming a globally averaged temperature and pressure at each level.  As also found by ZG12 and ZM13, the latitudinal temperature gradient is small.  In Cases 1 and 2, the temperature is warmer than Cases 3 and 4 because the bulk effect of a higher \methane~concentration.

The effect of initial surface pressure on atmospheric temperature is somewhat less straightforward.  Comparing Cases 1 and 2, it would seem that lower initial surface pressure corresponds to higher temperature at a given level.  Because \methane~is represented in the atmosphere by a percentage of the total atmospheric mass, a higher initial surface pressure, i.e. more massive atmosphere, means a larger amount of \methane~and in turn a warmer atmosphere.  The \methane~concentration is the same but the partial pressure (i.e. the number of molecules) of \methane~has increased, thus increasing the heating rate.  Yet, the opposite appears to be true.  This contradiction is an artefact of the way temperature has been plotted.  Had Cases 1 and 2 been plotted with pressure as the ordinate, it would be apparent that higher initial surface pressure corresponds to warmer temperatures at all altitudes.  At altitudes of $\sim$250 to 500 km, the temperature is almost isothermal, except for subtle effect of heating by the \methane~2.3 and 3.3~$\mu$m lines and cooling by the 7.6~$\mu$m line.  The temperature profile ``wobbles'' by about 10 km in this region, and the altitudes of local maxima and minima occur at different locations~\citep[see Fig.~8 of][for an example]{zalucha:2011a} depending on the initial surface pressure.

\subsection{Longitudinal Wind}\label{ss:u}

\begin{figure*}

 \noindent\includegraphics[width=1.0\textwidth]{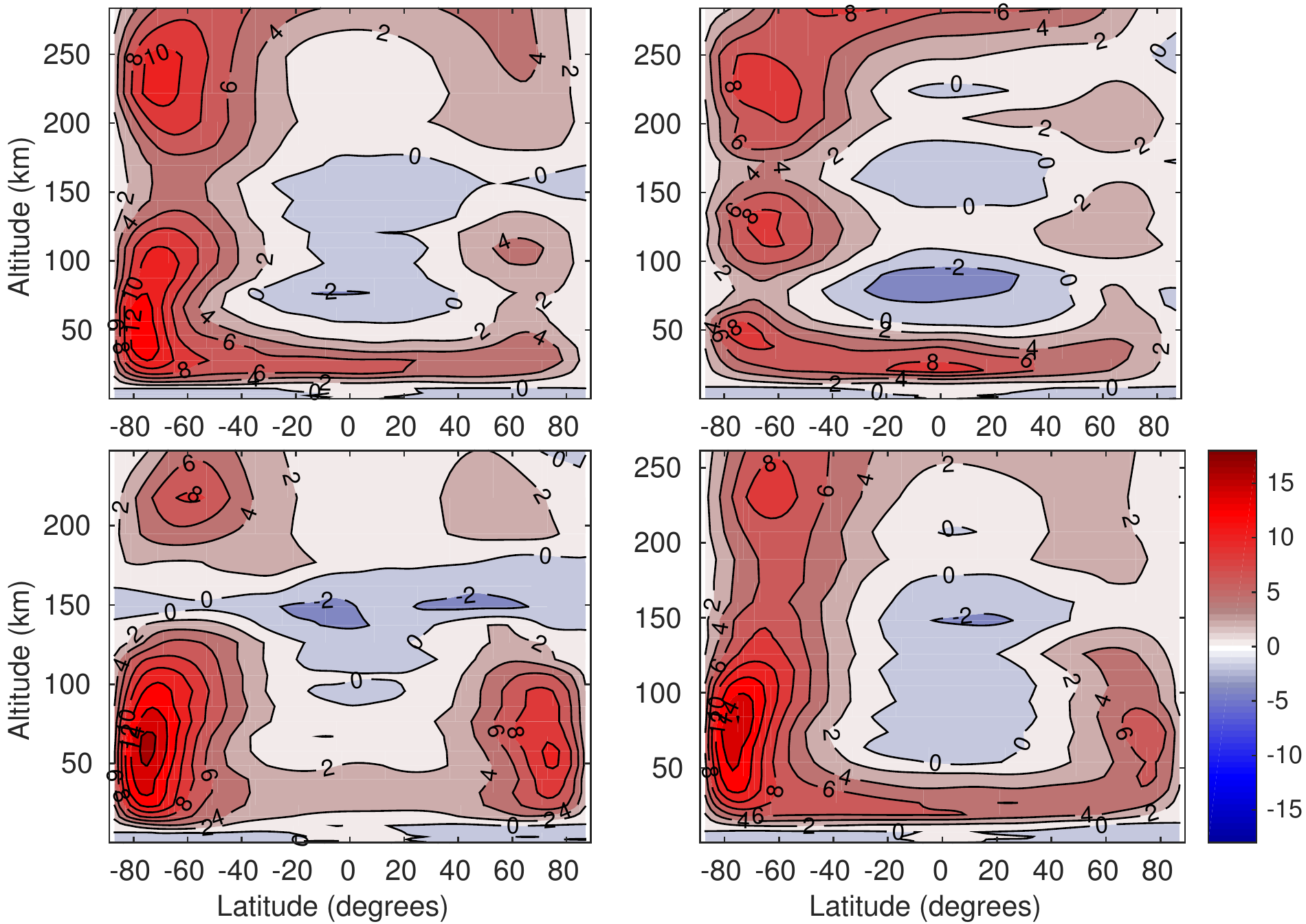}

 \caption{\label{fg:u1}  Longitudinally averaged westward (prograde) wind (\ms) for 2015 July 14.  Top left:  Case 1, top right: Case 2, bottom left: Case 3, bottom right: Case 4.  Positive values are westward (prograde).   For Case 1, the wind is strongest in the westward flow at $-75$\de~latitude between 50 and 100~km altitude.  Around 50~km altitude, the winds are westward at all latitudes, while above this layer there is an area of weak eastward winds at the equator.  For Case 2, the structure of the winds is the same as Case 1, except the maximum westward winds are weaker and the maximum eastward winds are stronger.  Case differs 3 the most from the other cases in that between roughly 200 and 250~km altitude, the winds are eastward everywhere.  The maximum eastward wind is also located between these altitudes, higher than the other cases.  Furthermore, the westward wind maximum is the strongest of the four cases.  For Case 4, the wind structure is similar to Case 1, except the westward maximum is slightly stronger, and the eastward maximum is located at a higher altitude.}
 
  \end{figure*}

In Fig.~\ref{fg:u1} the longitudinally averaged longitudinal wind (westward positive and prograde) is shown as a function of altitude and latitude for 2015 July 14.  In all cases, the wind in the lowest 10~km is light (less than 2~\ms).  This bulk measurement does not rule out small-scale phenomena such as plumes, dust devils, or flows from local topographic forcing; it is simply that the PGCM has a resolution of 50--60~km, which is too coarse to resolve such flows.  Just above 10~km altitude the winds are westward at all latitudes until about 60~km altitude where they change sign near the equator (except in the bottom left panel of Fig.~\ref{fg:u1}, where the sign change occurs at 100~km altitude).  This equatorial region of eastward winds extends from about $\pm$30\de latitude.  Note that this is approximately the width of Earth's tropics, but we believe this delineation is purely coincidental, since Pluto has a much different rotation rate, size, and horizontal temperature gradient compared with Earth.  The eastward winds are less than the westward winds, and the former do not exceed 4~\ms~in magnitude at any altitude.

In all cases, the winds are westward in the southern (i.e., winter) hemisphere south of $-30$\de latitude, with the exception of a narrow range of eastward winds around 150~km altitude for 8~$\mu$bar initial surface pressure and 0.2\% \methane~concentration (Case 3).  The maximum westward wind occurs between $-75$ and $-70$\de~latitude at an altitude of 50--70~km for \methane~concentrations of 0.2\%.  The maximum wind speed is of order 10~\ms.  The winds in the northern (i.e. summer) hemisphere north of 30\de~latitude are mostly westward as well, but slightly weaker.  In all cases there is a region of weak ($\sim$2~\ms) eastward wind aloft between $-20$ and 20\de latitude that varies in its altitudinal extent.  The altitudes and latitudes of local maxima and minima are mostly symmetric with the southern hemisphere.

Note that the structure of longitudinal winds does not at all resemble that of ZM13.  While the maximum westward and eastward winds (12~\ms and 2~\ms, respectively) are of the same order of magnitude, ZM13 found a westward maximum at the equator and 175~km altitude.  Between 50 and 175~km altitude there was a region of weak eastward winds northward of $-30$\de~latitude, and weak westward winds south of this latitude.  Below 50~km altitudes the winds were light and lacked structure with regards to direction.  The only similarity is that the maximum of the westward winds is of the same order of magnitude of the present study.  The differences between the model configurations are that ZM13 used the radiative-conductive scheme of~\citet{yelle:1989} (no CO or 2.3~$\mu$m \methane~line), the drag at the model top that was several orders of magnitude stronger and no \textbf{surface-atmosphere mass exchange scheme} was used. The effects of these processes on the circulation are examined as follows.

\begin{figure*}

 \noindent\includegraphics[width=1.0\textwidth]{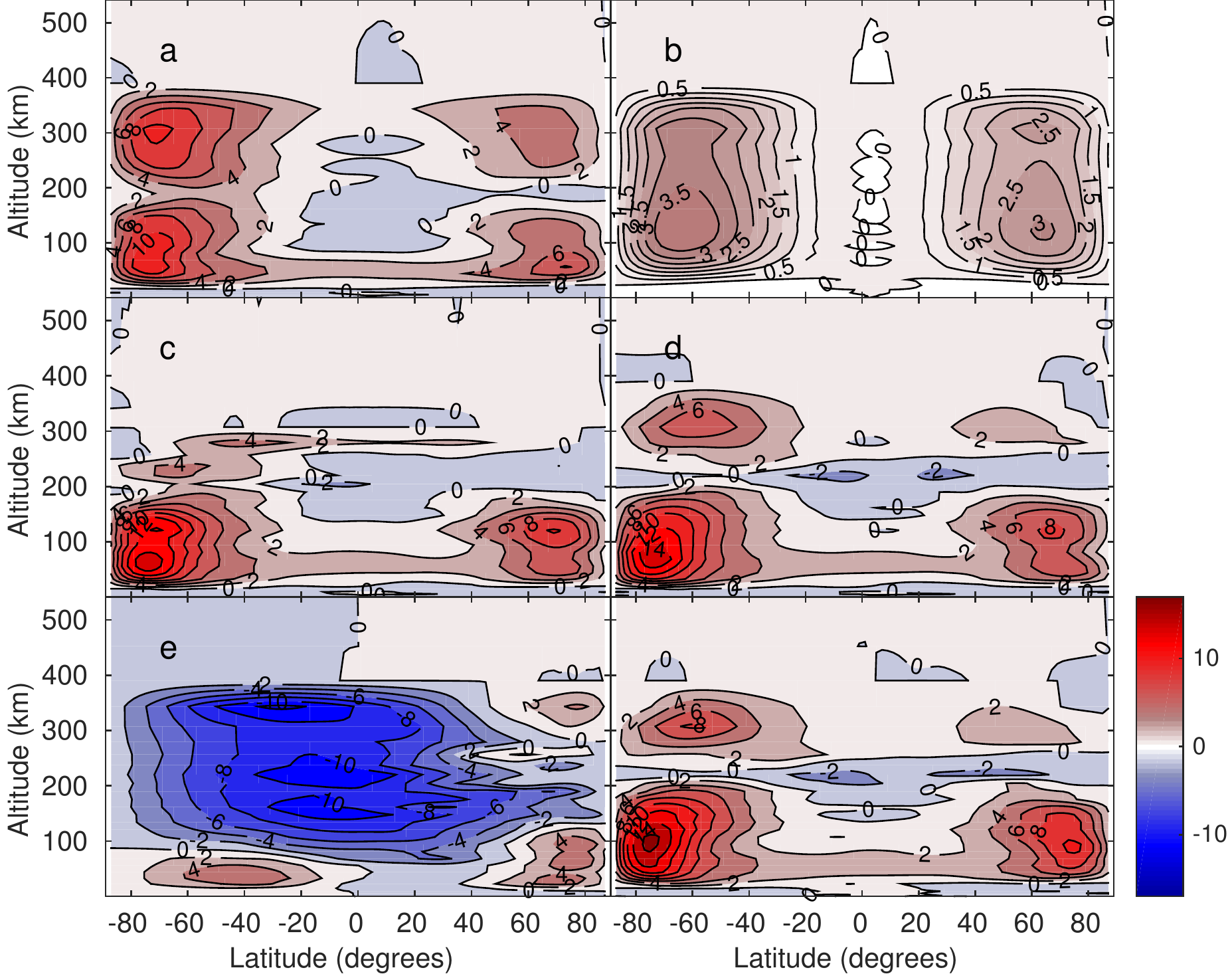}

 \caption{\label{fg:extraplots369}  Longitudinally averaged westward (prograde) wind (\ms) for Case 3 on 2015 July 14  Positive values are westward (prograde).  The panels are (a) no atmospheric CO cooling in the radiative-conductive scheme, (b) no contribution from the 2.3~$\mu$m \methane~line, (c) an upper atmospheric drag coefficient that is an order of magnitude higher than the operational runs and encompasses the top five layers instead of the top 3, (d) \textbf{no mass exchange between surface and atmosphere}, and (e) initialization at Northern Hemisphere winter solstice.  The panel with the adjacent colorbar is the same as Fig.~\ref{fg:u1}, bottom left panel for comparison.}
 
  \end{figure*}

Figure \ref{fg:extraplots369} shows the longitudinally averaged westward wind for Case 3, each with one of the model elements disabled, as well as an additional case run at Northern Hemisphere winter solstice and the full model results (hereinafter known as operational results) shown above for easy comparison.  The scenarios considered are: (a) no atmospheric CO cooling in the radiative-conductive scheme, (b) no contribution from the 2.3~$\mu$m \methane~line, (c) an upper atmospheric drag coefficient that is an order of magnitude higher than the operational runs and encompasses the top five layers instead of the top 3, (d) \textbf{no mass exchange between surface and atmosphere}, and (e) initialization at Northern Hemisphere winter solstice (integrated for the same amount of time as the operational runs).

The most striking difference is scenario (e), in which the winds are the opposite direction (i.e. retrograde or eastward) although of the same order of magnitude as the operational runs, maximum at the equator, and approximately symmetric at the equator.  Note the low-level area of westward winds close to the surface remains.  Until a follow up study is done with full Pluto-year simulations, it is not possible to know if this retrograde wind pattern is characteristic for this time or if there is a dependency on the model start time.  The latter is a property that \textbf{will certainly be investigated in the future. } 

Another difference from the operational run is apparent in scenario (c).  As expected, instead of the winds in the upper part of the domain going static around 300~km altitude, they now are damped down to altitudes between 200--250~km.  The circulation maxima near the poles are compacted; however note that the lowest level flow just above the surface is not impacted.  Unexpectedly, no drastic change in the circulation occurred due to purported waves reflecting off of the rigid lid caused by the substantially increased damping, except for in Case 3 where a thin band of eastward wind appears at about 225~km altitude over the equator and midlatitudes.

Scenario (b) is also clearly different in that the circulation splits into two distinct polar, westward jets with little to no flow at the equator and the disappearance of the near-surface westward flow that is present at all latitudes in the operational run.  Briefly note that in Scenario (a), where the CO cooling has been disabled, the circulation hardly differs from the operational run (except for the weakening of westward winds at the equator at 300~km altitudes).  Thus, Scenario (b) essentially recovers the case of \citet{yelle:1989}.  These authors had found that for the radiative-conductive equilibrium temperature, the 2.3~$\mu$m \methane~line was unimportant.  Later, \citet{strobel:1996} argued that is was important to the radiative-conductive temperature.  A further difference between these papers is the former assumes optically thin radiative transfer for near and thermal IR in the gray approximation for \methane~
bands and neglect of vibrational energy transfer from stretch modes to bending modes.  In any case, the results of Scenario (b) show that assumptions in the radiative-conductive scheme propagate into significant differences in the atmospheric circulation.

Scenario (b) is quite curious because except for \textbf{Case} 3, the polar structure is mainly preserved, but the westward winds intensify over the equator at altitudes between 200 and 250~km.  Again Case 3 stands out because an eastward area of wind develops over the equator.  Why the preference for Case 3 to develop these eastward jets is not easily inferred.  The near-surface westward winds remain, which calls into question whether these are related to the condensation flow.  Figures~\ref{fg:extraplots335}--\ref{fg:extraplots370} show the corresponding results for Cases 1, 2, and 4.

\begin{figure*}
 \noindent\includegraphics[width=0.75\textwidth]{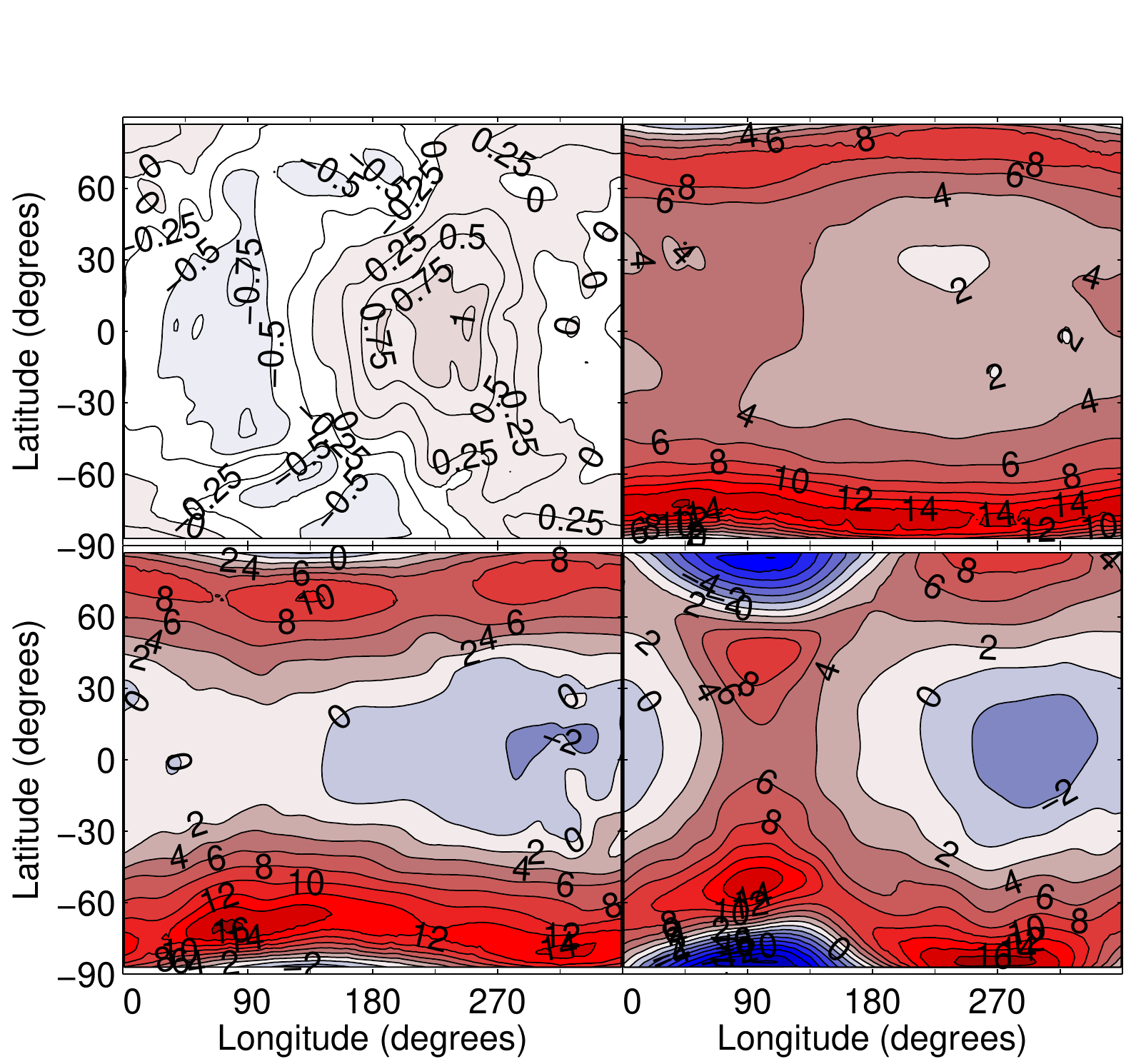}

\caption{\label{fg:uxy3}  Instantaneous westward (prograde) wind (\ms) for Case 3 on 2015 July 14.  Positive values are westward (prograde).  On this date at the time of the New Horizons flyby, the subsolar longitude was 32\de~and the subsolar latitude was 38\de~(in the pole convention used in this work, see text).  In the top left panel (2.2~km altitude), the magnitude of the maximum wind is of order 1~\ms, and the eastern (western) hemisphere is eastward (westward).  In the top right panel (32~km altitude), there is a strong flow (jet) of westward wind near the south pole, while elsewhere there is a local maxima of westward wind in the eastern hemisphere and a local minima in the western hemisphere.  In the lower left panel (98~km altitude) and lower right panel (248~km altitude), again there is a jet of westward wind near the south pole.  There is a weaker flow of westward winds near the north pole (and a localized area of eastward wind very close to the north and south poles).  At 217~km altitude there is an area of weak westward flow in the eastern hemisphere.}

  \end{figure*}

Figure~\ref{fg:uxy3} show horizontals cross-sections of longitudinal wind at key altitudes for Case 3 on 2015 July 14.  Near the surface at 2.2~km altitude (upper left panels), the wind is weak, but there is a distinct wave number-1 pattern between westward and eastward directions.  At the next altitude shown (32~km) there is a strong westward jet that is of constant strength with longitude.  Elsewhere (except very near the north pole), the winds are also westward, but there is a local minima near 270\de~longitude and a local maxima near 90\de.  At 98~km altitude, the winds are weakly eastward between $-30$\de~and 30\de~latitude (except for an area of westward winds in the 24~$\mu$bar, 1\% \methane~case) with a slight variation with longitude.  The winds are relatively strongly westward near the poles.  Finally, at 217~km altitude, there is a wave number-1 structure between $-60$\de~and 60\de~latitude, with a local maximum of eastward winds at 270\de~longitude and a local maximum of westward winds at 90\de~longitude.  Poleward of these latitudes, the location of the eastward and westward maxima are shifted by 180\de.  Figures~\ref{fg:uxy1}--\ref{fg:uxy4} show the corresponding results for Cases 1, 2, and 4.

\subsection{Latitudinal and Vertical Wind}
\begin{figure*}
 \noindent\includegraphics[width=0.75\textwidth]{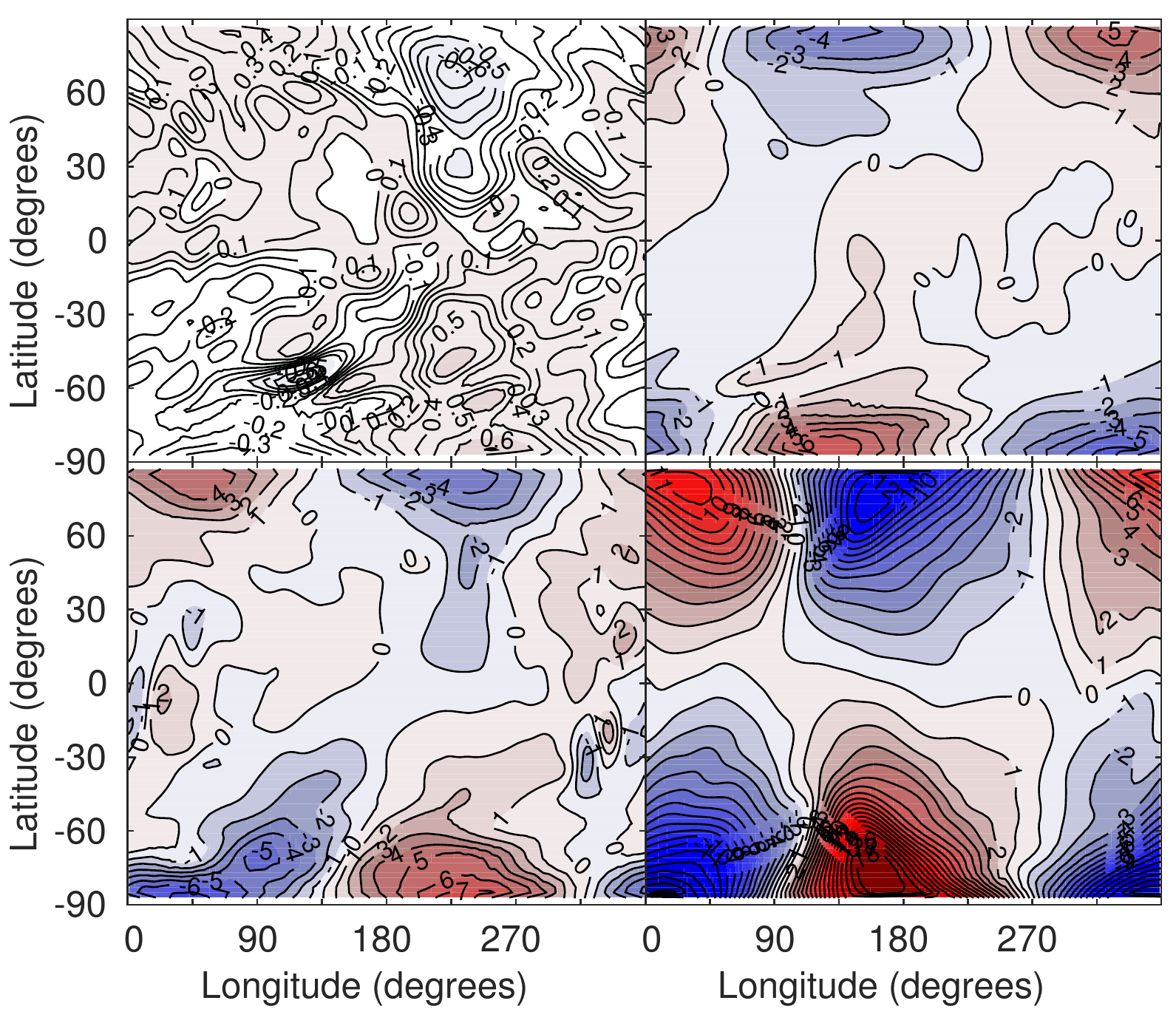}

\caption{\label{fg:vxy3}  Instantaneous northward wind (\ms) for Case 3 on 2015 July 14.  Positive values are northward.  On this date at the time of the New Horizons flyby, the subsolar longitude was 32\de~and the subsolar latitude was 38\de~(in the pole convention used in this work, see text).   Top left panel: 2.2~km altitude, top right panel: 32~km altitude, bottom left panel: 98~km altitude, bottom right panel: 217~km altitude. On this date, the subsolar point is located at 32\de~longitude.  Except at the lowest level, air appears to be flowing over the poles along the terminators.  The magnitude of the maximum flow is about 10\ms.}

  \end{figure*}

In the longitudinal average, the latitudinal and vertical wind do not show any structure.  However, when plotted in the xy-plane, at the same altitudes as Fig.~\ref{fg:uxy3}, each show distinct variations.  Figure~\ref{fg:vxy3} shows the latitudinal wind for Case 3 (Figs.~\ref{fg:vxy1}, \ref{fg:vxy2}, and \ref{fg:vxy4} show the corresponding results for Cases 1, 2, and 4).  At the equator, the latitudinal flow is nearly motionless.  At low altitudes, again the flow is weak for the same reason it was in the longitudinal direction.  There are definitive areas of northward (positive) and southward (negative) flow near the poles (poleward of 60\de~latitude).  For the uppermost three altitudes shown, between approximately 90\de~and 270\de~longitude the flow is southward at the north pole and northward at the south pole; i.e. away from the pole.  At other longitudes, the flow is towards the pole.  For a given longitude, the flow direction is opposite that at the opposite pole (e.g., where there is northward flow at the south pole, there is southward flow at the north pole and vice-versa). Remembering that we are on a 3-D sphere, these figures suggest that air is flowing over the poles.

When all 30 levels of the model are examined (not shown), the areas of northward and southward flow at the poles shift eastward with altitude.  Thus, the air always flows over the poles (except near the surface), but the meridian it follows shifts eastward with height.  In Fig.~\ref{fg:uxy3}, the altitudes chosen all show northward and southward flow confined to the same longitudes.  Looking at Case 1 (Fig.~\ref{fg:uxy1}, the top two altitudes plotted (bottom two panels) show the areas of maximum northward and southward flow 180\de~out of phase in longitude.  That is, in the lower left panel between longitudes 90\de~and 225\de~the flow is southward at the south pole, but in the lower right panel the flow is northward at these same longitudes.  \textbf{The reasons that the areas of maximum north and south flow shift eastward with height is not yet understood}.

\begin{figure*}
 \noindent\includegraphics[width=0.75\textwidth]{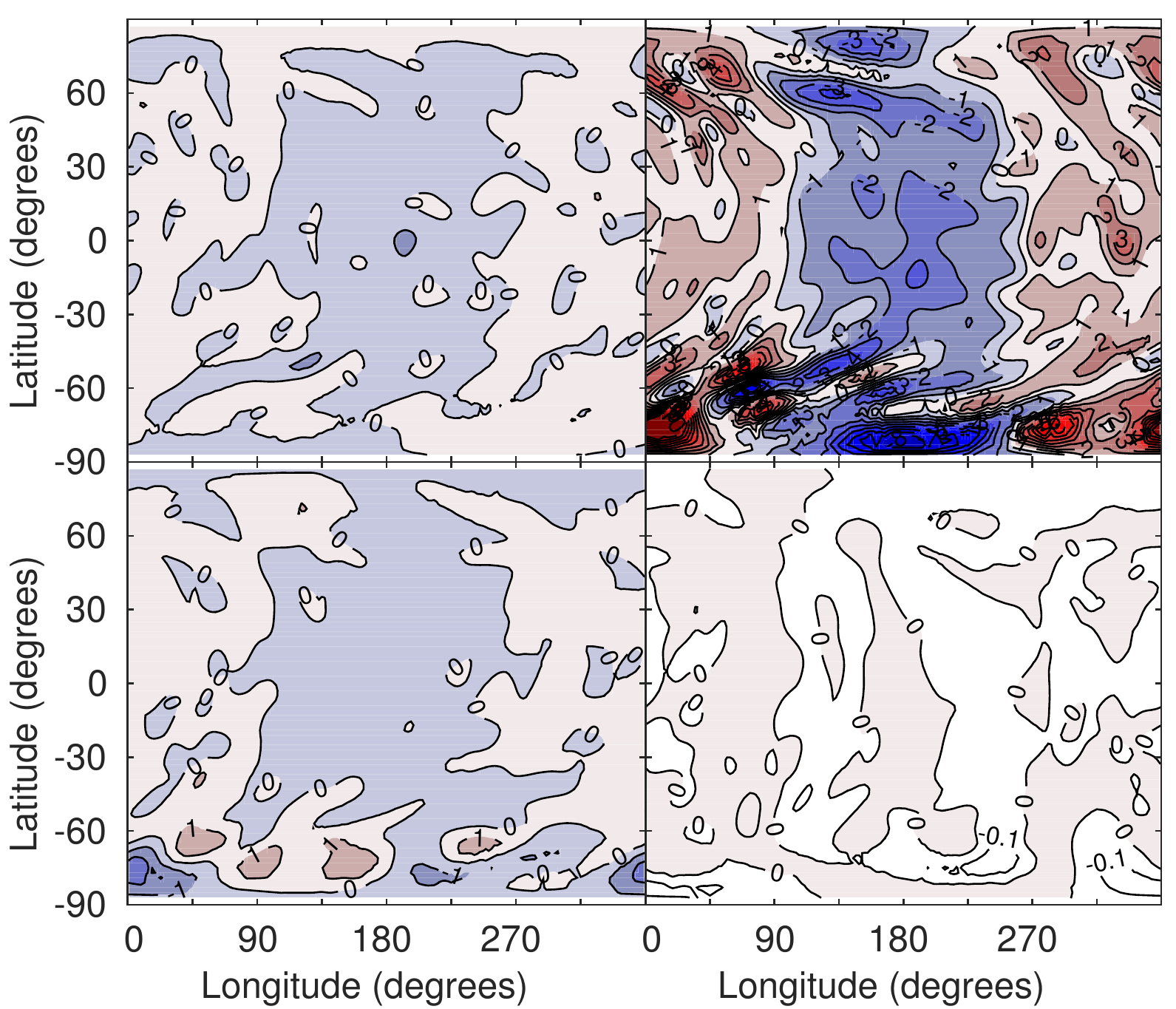}

\caption{\label{fg:wxy3}  Instantaneous vertical wind ($10^{-6}\mu$bar~s$^{-1}$) for Case 3 on 2015 July 14.  Positive values are downward.  On this date at the time of the New Horizons flyby, the subsolar longitude was 32\de~and the subsolar latitude was 38\de~(in the pole convention used in this work, see text).  Top left panel: 2.2~km altitude, top right panel: 32~km altitude, bottom left panel: 98~km altitude, bottom right panel: 217~km altitude. On this date, the subsolar point is located at 32\de~longitude.  Except at the lowest level, air appears to be flowing over the poles along the terminators.  The magnitude of the maximum flow is about 10\ms.}

  \end{figure*}

Figure~\ref{fg:wxy3} shows the vertical wind in pressure coordinates for Case 3.  In this \textbf{coordinate} system, positive is downwards, negative is upwards, and the units are in $10^{-6}\mu$bar~s$^{-1}$.  There is a clear area of up-welling between 90\de~and 270\de~longitude, with down-welling at other longitudes, except at the highest altitude shown, where there is little vertical motion.  There is also fine-scale structure that is almost as large in magnitude as the maxima and minima.  What we may conclude from the meridional and vertical wind directions is that the flow on Pluto is not like other bodies in the solar system that show overturning circulations in the longitudinal mean, i.e. Hadley cells.  Future work explaining the unusual atmospheric dynamics is needed.  Figures~\ref{fg:wxy1}--\ref{fg:wxy4} show the corresponding results for Cases 1, 2, and 4.

\section{Model Stellar Occultation Light Curves}\label{se:lc}
Stellar occultation light curves produce a data product in the form of number of photon counts (flux) vs. time~\citep[see][for a review]{elliot:1996}.  To turn these data into a form that is more physically intuitive, the data are fit or inverted to obtain temperature vs. altitude (or radius from Pluto's center).  This inversion technique can induce large errors between the inverted temperature profile and the true temperature profile (as tested with synthetic light curves).  Model fits assume a certain structure.  Thus, when comparing light curve models, it is best to take a temperature profile and carry out the \textit{forward} problem---calculate a model light curve and compare data and model in the flux vs. time (or flux vs. altitude) regime.

\citet{chamberlain:1997} and \citet{zalucha:2011a} describe in detail how to calculate a model light curve.  Here the procedure is briefly summarized.  Given a temperature profile as a function of pressure (in this Case from the PGCM), we may use the ideal gas law and the law of hydrostatic balance to obtain number density as a function of distance from Pluto's center.  The number density is proportional to the refractivity, and using the laws of geometric optics, we may obtain the bending angle and its derivative with radius.  Once these quantities are known, along with the distance from the observer to the occulting body, a light curve may be calculated.

\begin{figure}

 \noindent\includegraphics[width=1.0\columnwidth]{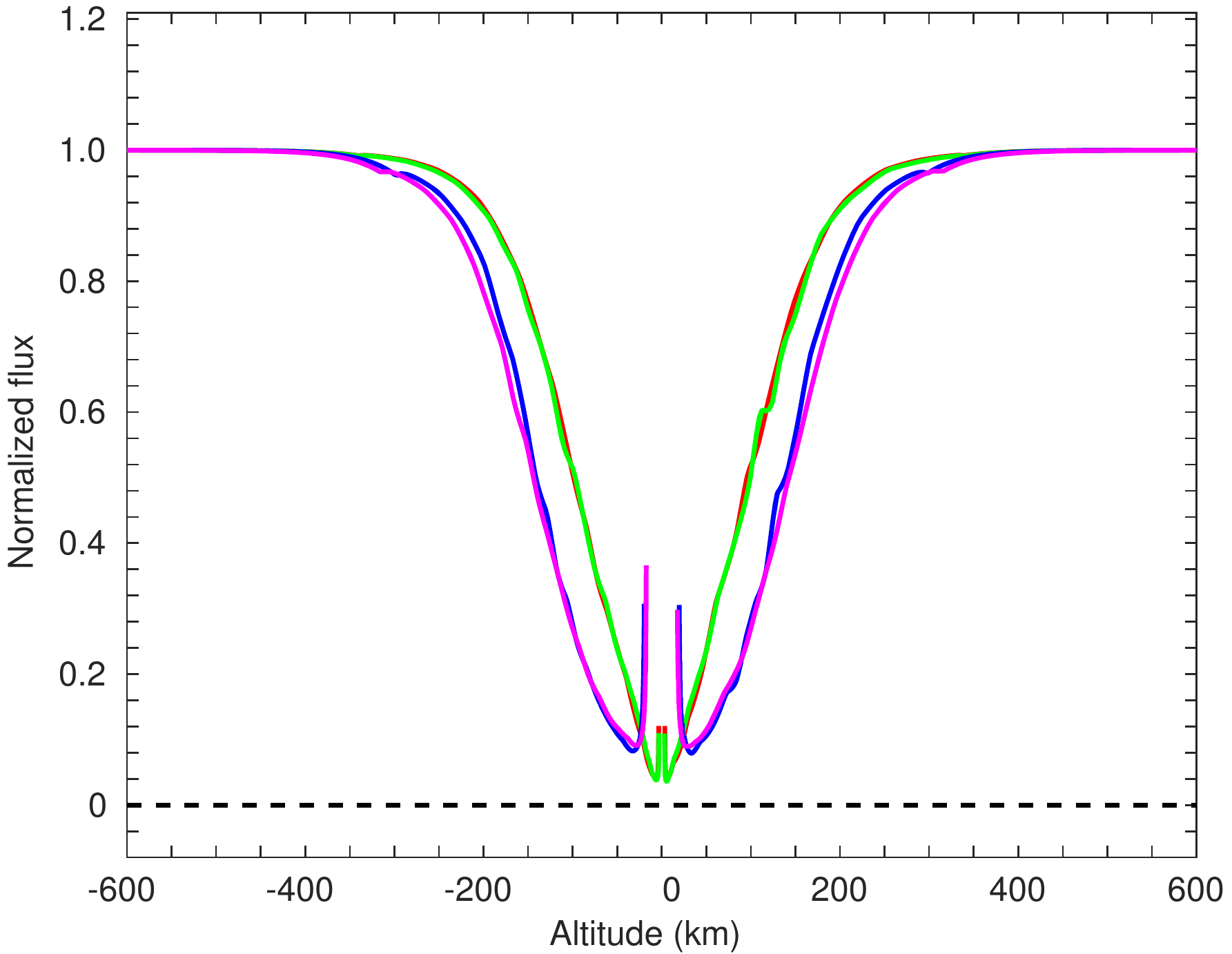}

 \caption{\label{fg:lc}  Model stellar occultation light curves. Red: initial surface pressure 8~$\mu$bar and \methane~concentration 1\%, green: initial surface pressure 24~$\mu$bar and \methane~concentration 1\%, dark blue: initial surface pressure 8~$\mu$bar and \methane~concentration 0.2\%, magenta: initial surface pressure 24~$\mu$bar and \methane~concentration 0.2\%.  The red and green curves are almost indistinguishable as are the blue and magenta curves.  A ground-based observer could distinguish between \methane~concentrations of 0.2\% and 1\% using the PGCM, but probably not between pressures of 8 and 24~$\mu$bar.
}

  \end{figure}

Figure~\ref{fg:lc} shows an example of what ground-based observers would observe in the year 2015 given different atmospheric configurations.  Extinction by aerosols or other particles has not been taken into account, because including them would be inconsistent with the PGCM not accounting for their radiative effects; thus, an upturn (i.e., the central flash due to diffraction around Pluto's disk) is present near the middle of the light curve.  The light curve has been cast as intensity vs. altitude in Pluto's atmosphere.  To convert a particular observing geometry and the quantities that are measured, i.e. flux vs. time, one uses the equations:
\begin{eqnarray}
s(r)&=&r+D\theta(r)\\
t&=&t_{mid}\pm\sqrt{\frac{s(r)^2-s_{mid}^2}{V^2}},
\end{eqnarray}
where $s$ is the observer's position in the shadow (observer's) plane, $r=z+r_S$ (where $r_s$ is the body's radius) is the distance from the body's center in the body (Pluto) plane, $D$ is the distance between the observer and the body, $\theta$ is the bending angle, $t_{mid}$ is the midtime of the occultation, $s_{mid}$ is the distance of closest approach (i.e., impact parameter) in the shadow plane, and $V$ is the relative velocity between the observer and the body.  In Fig.~\ref{fg:lc}, there is a clear difference between the light curves derived from PGCM simulations with different \methane~concentrations and a slight difference between light curves derived from simulations with different initial surface pressures.  The shape of a light curve depends on the temperature gradient with height in the atmosphere~\citep{elliot:1992}.  The behaviour of the light curves is consistent with the PGCM temperature results (Fig.~\ref{fg:t1}), where there is a larger difference between the cases with different \methane~concentrations than there is between different initial surface pressures.  This behaviour of the light curves is opposite that of \citet{zalucha:2011a}, who found that when using the steady state version of the~\citet{strobel:1996} model, the greatest difference appeared in light curves of different surface pressure and the difference between \methane~concentrations was secondary.  

There is some difference in how the \citet{strobel:1996} was applied in \citet{zalucha:2011a} and this work.  First, the original  \citet{strobel:1996} version assumed disk averaged radiation, while in the process of implementing it into the PGCM, dependence on insolation in the column as a function of latitude, longitude, and season was added.  Second, when the temperature is calculated in the PGCM, it is subject to not only the heating and cooling terms of \citet{strobel:1996}, but also other temperature modifying terms: advection, adiabatic heating and cooling, and numerical filters.  These in turn interplay with the entire system of equations that comprise the PGCM.  It is certain that the \citet{strobel:1996} model has been correctly implemented into the PGCM because if the calls to the dynamic subroutines and other subroutines are disabled, the results of the stand-alone \citet{strobel:1996} model are recovered.  In summary, a ground-based observer would be able to distinguish between \methane~concentrations of 0.2\% and 1\% using the PGCM, but probably not between pressures of 8 and 24~$\mu$bar.

\section{Predictions for New Horizons}\label{se:nh}
When a radio occultation experiment is performed, radio waves of a known frequency are transmitted through an atmosphere and are measured by a receiver.  The rays bend as they travel through the atmosphere due to diffraction of light in a medium.  Unlike shorter wavelengths of light, the differential refraction (i.e. flux difference) of radio waves due to the density gradient in an atmosphere is too small to be measured.  What is measured is the difference in Doppler-shifted frequencies due to light travelling through the atmosphere compared with empty space.  The bending angle can be calculated given precise knowledge of the geometry of the transmitter and receiver~\citep{hinson:1999} in the limit of geometric optics.  The bending angle as a function of impact parameter relative to the occulting body's center may then be transformed into the refractive index as a function of radius from the occulting body's center.  The index of refraction, rather the refractivity, is proportional to the number density, and using the ideal gas law and the assumption of hydrostatic balance, temperature as a function of height may be obtained.  The solar and stellar occultation experiment performed by Alice use the same equations as in Section~\ref{se:lc}.

\begin{figure}

 \noindent\includegraphics[width=1.0\columnwidth]{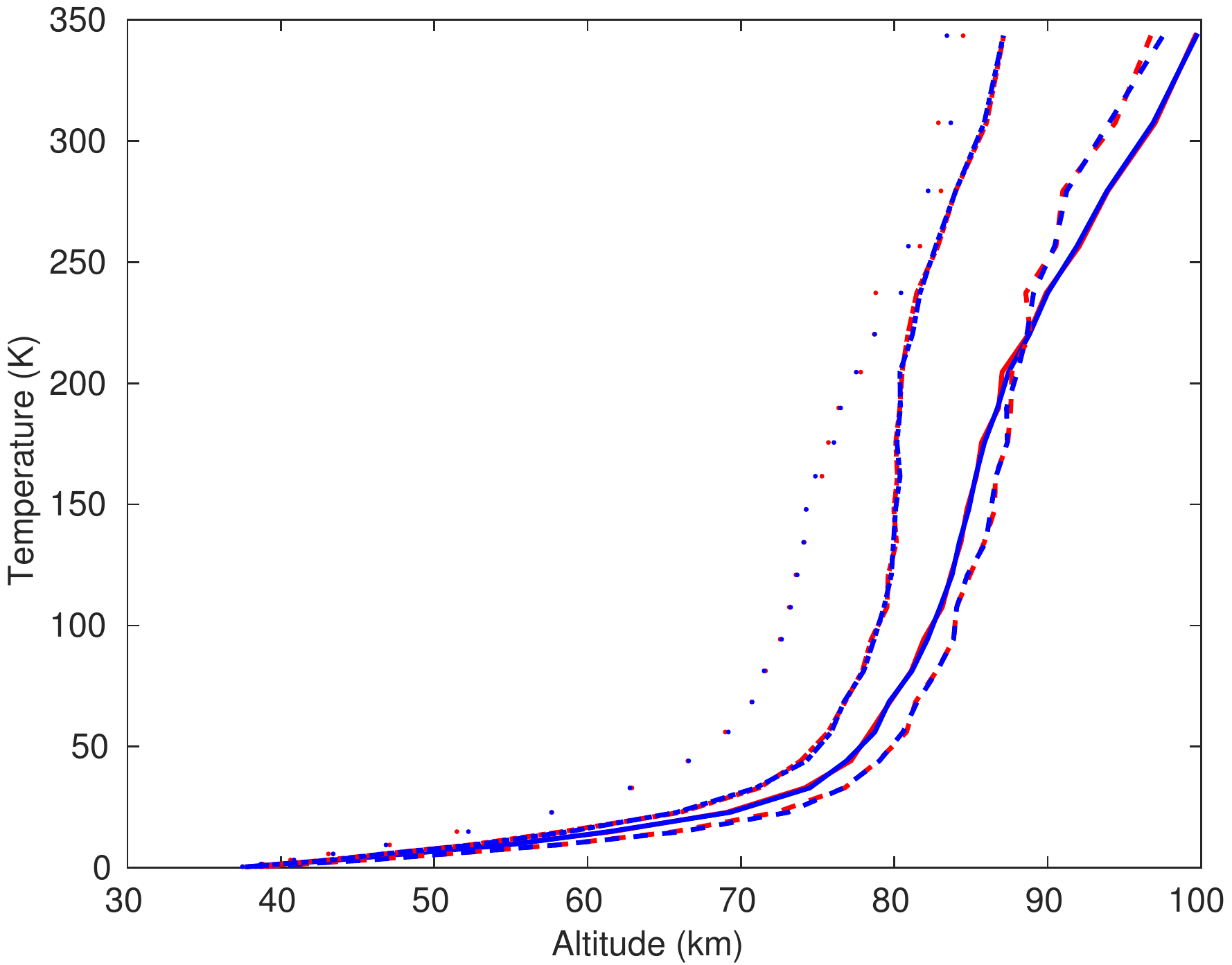}

 \caption{\label{fg:alice_sun_t}  Temperature profiles predicted to be derived from Alice observations during the solar occultation.  Red: immersion, blue: emersion.  Solid line: initial surface pressure 8~$\mu$bar and \methane~concentration 1\%, dashed line: initial surface pressure 24~$\mu$bar and \methane~concentration 1\%, dotted line: initial surface pressure 8~$\mu$bar and \methane~concentration 0.2\%, dot-dashed line: initial surface pressure 24~$\mu$bar and \methane~concentration 0.2\%.  The immersion and emersion profiles do not differ greatly, which is to be expected since the horizontal temperature gradients are weak.  The temperature profiles for different initial surface pressures and \methane~concentrations differ by up to 20~K in some places, which will be detectable by New Horizons.}

  \end{figure}

\begin{figure}

 \noindent\includegraphics[width=1.0\columnwidth]{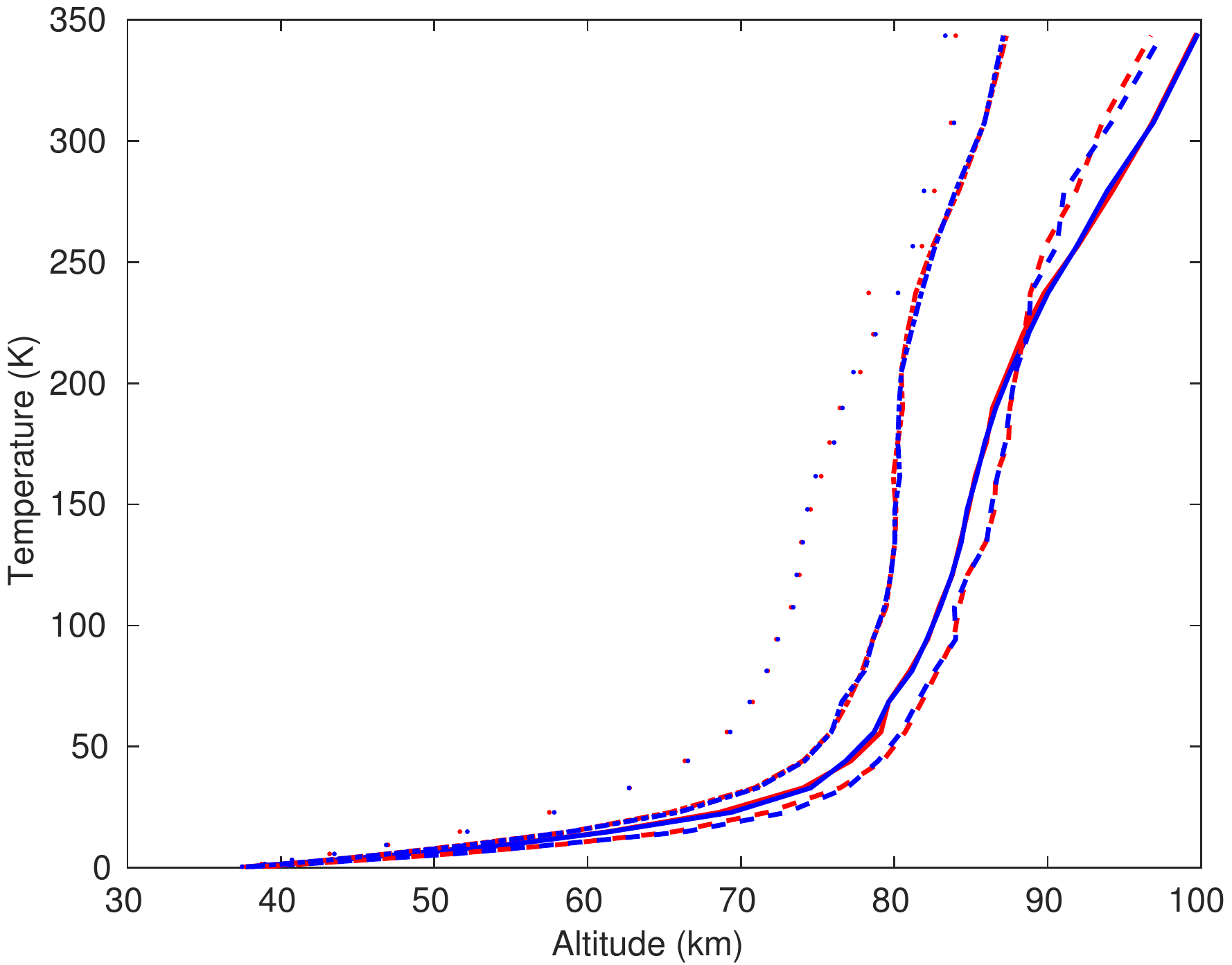}

 \caption{\label{fg:alce_star_t}  Same as Fig.~\ref{fg:alice_sun_t}, but for the Alice stellar occultation.}

  \end{figure}

\begin{figure}

 \noindent\includegraphics[width=1.0\columnwidth]{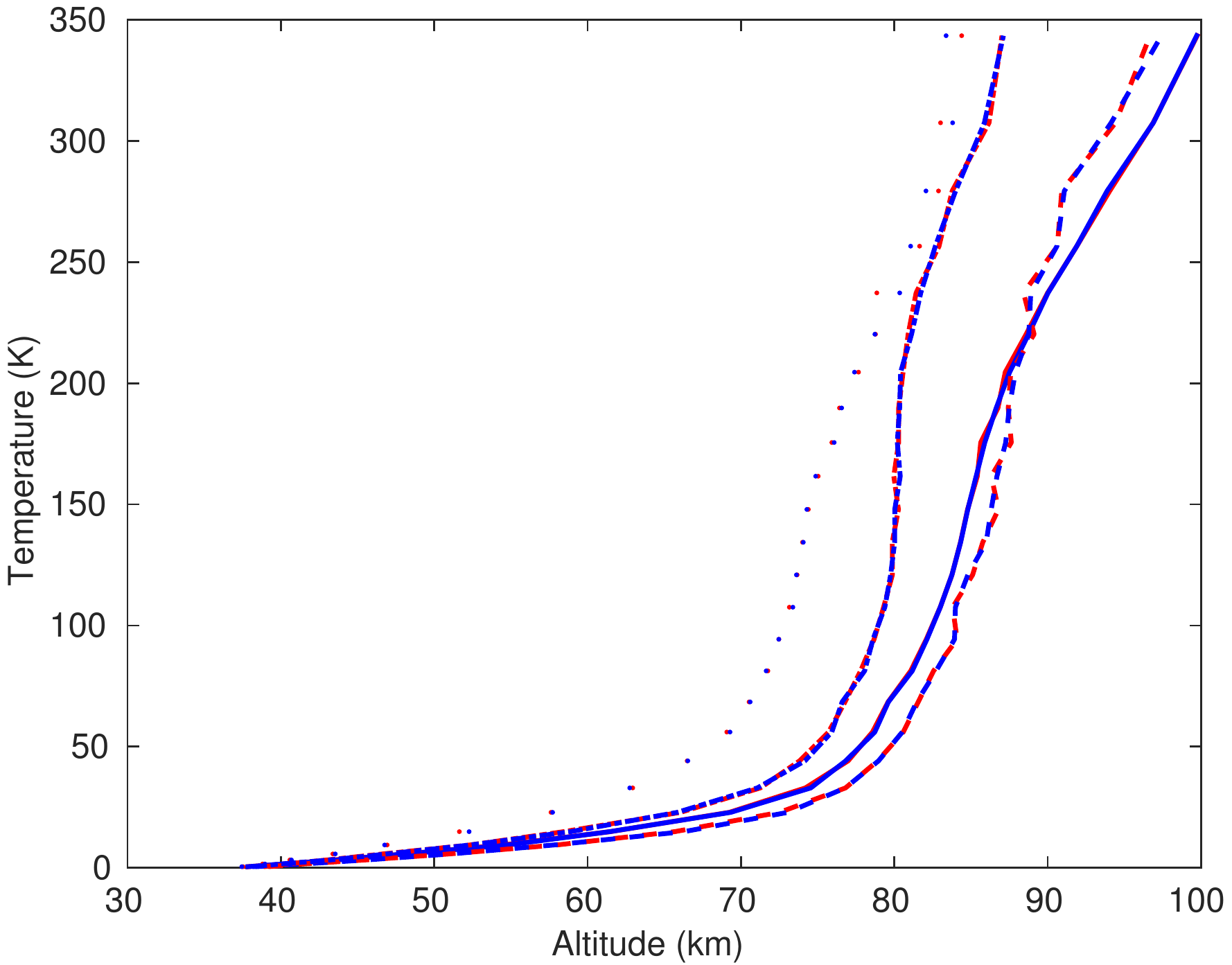}

 \caption{\label{fg:REX_t}  Same as Fig.~\ref{fg:alice_sun_t}, but for REX the radio occultation.}

  \end{figure}

Table~\ref{tb:nhcoords} shows the longitudes and latitudes that each of the New Horizons occultation experiments probed.  Figs.~\ref{fg:alice_sun_t}--\ref{fg:REX_t} show the temperature profiles predicted to be derived from by Alice (sun), Alice (background star), and REX observations.  The Alice and Rex profile altitudes were calculated with the temperature and pressure of that column.  There is little difference between the immersion and emersion profiles and the different experiments, owing to the weak horizontal temperature gradients described in Section~\ref{ss:pgcm_t}.  The effect of \methane~concentration may be seen (c.f. solid and dotted, dashed and dot-dashed lined) and initial surface pressure (c.f. solid and dashed, and dotted and dot-dashed).  Near 25~km altitude, the temperature profiles differ by as much as 20~K, which will be distinguishable by New Horizons.  Above 50~km altitude, the maximum difference is as much as 10~K, which is still distinguishable, but in some places the difference is zero.  However, as we are comparing the whole temperature profile and not individual points, the PGCM should be able to tell the difference between different initial surface pressures and \methane~concentrations.  Further simulations may be needed to pinpoint the combination of initial surface pressure and \methane~concentration, which can be carried out if necessary.

\begin{table}
\centering
\caption{Longitude and latitude coordinates for New Horizons occultations\label{tb:nhcoords}}
\begin{tabular}{lrrr}
\hline
Coordinate & Alice & Alice~~~~~~~~~ & REX \\
           & (Sun) & (Background star) & \\
\hline
Immersion longitude & 150\de & 170\de~~~~~~~~~ & 164\de \\
Immersion latitude & 20\de & 40\de~~~~~~~~~ & 15\de  \\
Emersion longitude &350\de & 340\de~~~~~~~~~ & 343\de \\
Emersion latitude & $-20$\de & $-10$\de~~~~~~~~~ & $-14$\de \\
\hline
\end{tabular}

\end{table}

\section{Discussion}
\citet{toigo:2015} performed simultaneous, independent simulations of Pluto's atmosphere from this work. They used a different approach, by which they ran the surface model to equilibrium before coupling it to a GCM.  The dynamical core is that of planetWRF~\citep{richardson:2007}.  They \textbf{used} the \citet{zhu:2014} scheme (the successor to \citet{strobel:1996}, mainly adjusted to predict conditions at very high altitudes) for atmospheric heating and cooling.  There are some fundamental differences between the dynamical cores of planetWRF and the MIT GCM.  First, planetWRF uses a finite difference scheme while the MIT GCM dynamical core uses a finite volume scheme.  Numerically, the former approximates derivatives as discrete differences between grid points, whereas the latter casts the primitive equations of fluid dynamics in terms of fluxes, and calculates fluxes across grid box boundaries.  A second difference between planetWRF and the MIT GCM dynamical cores is that planetWRF uses a latitude longitude grid in spherical coordinates, while the MIT GCM uses a cubed-sphere grid: a projection of a cube onto a sphere.  The drawback of a spherical projection is that near the poles, the grid points become very close to each other, eventually turning into a numerical singularity at the pole.  In order to not violate the CFL criterion, a prohibitively small time step is necessary.  To get around this issue, users of spherical grids use a numerical filter to artificially dampen small scale oscillations.

The results of this work and \citet{toigo:2015} agree in that the latitudinal temperature gradient is weak.  However due to the fact that \citet{toigo:2015} uses much different boundary conditions, which among themselves are highly different~\citep[following that of][]{young:2013}, they obtain different wind structures.  Because at present \textbf{a coupled GCM and subsurface/surface model has been spun-up simultaneously}, it cannot be said which initialization method for the lower boundary condition, i.e that of this work or those of \citet{toigo:2015}, is the proper one (though in light of New Horizons findings, both will need to be modified).  It is not known if equilibrating the surface frost with a minimally complex atmosphere is adequate or if the subsurface/surface and atmosphere are tightly coupled.  Nor is it known that if they are tightly coupled, what the implications are for the surface frost distribution and atmospheric state variables.  At this point in time we may only describe the effects of the atmosphere given a certain boundary condition and compare with observables such as temperature and surface frost distribution.  This issue will only be resolved when a fully coupled PGCM \textbf{is} integrated until it has reached equilibrium (determined by repeatability of seasonal surface frost and quasi-equilibrium of globally averaged atmospheric kinetic energy).

\section{Conclusions}\label{se:conclusions}
This paper has presented results from a PGCM based on the MIT GCM.  It has continued the work of ZG12 and ZM13, who also used the MIT GCM as a basis.  This latest version uses the radiative-conductive scheme of~\citet{strobel:1996} for atmospheric heating and cooling, which adds cooling by CO and heating by the 2.3~$\mu$m \methane~line to the \citet{yelle:1989} model used in previous versions; \textbf{mass exchange between the surface and atmosphere}; a \textbf{surface and} subsurface model; a drag force (to prevent reflections off the artificial model top and represent buoyancy waves) at the model top that is several orders of magnitude weaker than previous versions; and a spin up procedure that has Newtonian relaxation step at the beginning to more gradually initialize the model, which allows for a larger time step (and thus a more efficient use of super-computing time). \textbf{ This model of Pluto is on}e of two very similar models developed at the same time by independent groups~\citep[the other being][]{toigo:2015}. 

\textbf{Importantly}, the PGCM can predict wind, a quantity that is difficult to measure remotely even for a satellite in orbit or spacecraft flybys.  Special circumstances must be present, for example relating ocean wave heights to surface wind (Earth and Titan), observing tracers such as clouds or plumes (Venus, Earth, Triton, and Jupiter), or observing wind surface patterns in dunes or ice (Mars and Triton).  Often these phenomena are transient or only provide wind direction and not speed.  The fact that the PGCM can predict winds over the entire globe at any time is a key power shared with GCMs in general.

The model was integrated for 30 Earth-years (0.12-Pluto years)from the years 1985 to 2015 to cover the stellar occultation observational record.  Due to the computationally prohibitively expensive spin-up time to equilibrate the atmosphere and the surface, it is unknown if the model achieved equilibrium.  But, based on other modelling and observational studies it is not likely the case.  Thus, four simulations were performed that varied initial global mean surface pressure and methane concentration.  As in ZM13, there is effectively no longitudinal (diurnal) \textbf{atmospheric} temperature variation, and the latitudinal \textbf{atmospheric} temperature gradient is small.  The first property follows from the estimation that the radiative-conductive time-scale is much longer than a Pluto day.  However it has yet to be explained why a large latitudinal gradient is not predicted on Pluto. Given the non-uniform ice distribution and composition observed by New Horizons~\citep{stern:2015}, this property is likely to change. 

The longitudinal wind has a completely different structure than ZM13 (and ZM12, but this is to be expected since that model was 2-D), but the maximum magnitude of order 10~\ms~is the same.  In the current study, the longitudinal winds are of order 1~\ms~below 10~km.  There is a wave number-1 structure in this altitude zone in the longitudinal direction.  Above 10~km altitude upwards to 75~km altitude (except in one case where this region extends to 130~km altitude), the wind is westward everywhere.  Near the south pole there is a strong westward jet, representing the global maximum wind speed.  Above this region up to 175~km altitude, there is a region of weak ($\sim$1~\ms) eastward winds between $\pm30$\de~latitude.  Poleward of this latitude the winds are westward.  From 175~km altitude to the base of the drag layer (350~km altitude) there is a wave number-1 structure between $\pm$60\de~latitude with a local maximum of eastward winds at 270\de~longitude and a local maximum of westward winds at 90\de~longitude.  Poleward of these latitudes, the location of the eastward and westward maxima are shifted by 180\de.

In the longitudinal average, the latitudinal and vertical winds averaged to approximately zero.  However when viewed on horizontal planes at different altitudes, it is apparent that air is flowing over the poles (albeit along different meridians depending on height), with broad areas of up-welling east of the subsolar point and down-welling west of the subsolar point.  Once again we find Pluto's atmosphere is \textbf{unlike} any other in the solar system.  There is a caveat that the subsurface/surface and atmosphere may not have had time to equilibrate, due to the computationally prohibitive wall clock times to run a simulation for even only one Pluto year.  Thus it is possible the flow will be different, if and when it is possible to carry out such long-term simulations.

Some notable observations were made when simulations were repeated with various processes left out.  First, when the model was initialized at Northern Hemisphere winter solstice, the longitudinal winds became eastward almost everywhere with a maximum above the equator.  When the contribution from the 2.3~$\mu$m \methane~band was removed, the winds became stagnant at the equator.  When the upper atmosphere drag force was increased and extended to lower levels, the top of the longitudinal flow lowered, but the structure stayed the same, only compressed.  Little difference was seen when the effects of CO were removed from the radiative-conductive scheme.  When the \textbf{surface-atmosphere mass exchange was turned off}, the longitudinal winds just below the damping layer strengthened at the equator.

Predictions for the temperature that will be derived from New Horizons Alice and Rex observations were provided.  The location (i.e. latitude and longitude) that each instrument sampled did not produce significant difference in the temperature profiles, since temperature gradients in the PGCM results were weak.  This includes immersion and emersion profiles.  In the stratosphere, the temperature profiles differ by as much as 20~K, which will be distinguishable by New Horizons.  In the mesosphere (the more isothermal layer above 100~km) the maximum difference is as much as 10~K, which is still distinguishable, but in some places the difference is zero.  However, as we are comparing the whole temperature profile and not individual points, the PGCM should be able to tell the difference between different initial surface pressures and \methane~concentrations.

In the predictions for ground-based stellar occultations, the difference between light curves of different \methane~concentrations would be detectable, but the difference between light curves of different initial surface pressures would not.  While this result is opposite that of \citet{zalucha:2011a}, we can rule out coding error by de-activating everything but the call to the \citet{strobel:1996} module.  The difference must lie in the fact that the \citet{strobel:1996} module of the PGCM has longitudinally, latitudinally, and seasonally varying insolation, and that the model is no longer in the steady state ($dT/dt=0$), but interplay with the rest of the model (Navier stokes equations, mass and energy equations, ideal gas law, numerical filters).

\textbf{Future simulations will test the subsurface and surface models and the volatile cycle before conducting runs spanning multiple Pluto years.  It is speculated that this }is enough time to spin up the volatile cycle, i.e. the location and depth of surface ice, rather than assume effectively infinite ice cover.  Mars GCMs require about two Martian years to equilibrate the ice cycle.  From~\citet{hansen:1996}, it was found that the frost cycle was spun up after four Pluto years.  This model had a one-layer atmosphere, so it is unknown at this time what the effect of a complete surface and atmosphere model will require for spin up.  A fully spun up volatile cycle might affect the results for winds and temperature, particularly the latitudinal wind, as air flows from the sublimating summer pole (an area of high pressure) to the condensing winter pole (an area of low pressure), as is seen in in the lowest atmospheric layer of this PGCM and Mars GCMs. 

An interesting insight into volatile transport in the PGCM would be to use the ``tracer'' package. This feature allows the user to initialize inert particles in the atmosphere, and the scheme keeps track of where they are transported (though making these particles radiatively active would not be computationally difficult).  This treatment would allow us to also keep track of sublimated N$_2$ (including if it froze and resublimiated) if say it was transported up to another layer.  Future work is planned to carry out these simulations.



\section*{Acknowledgements}
This work used the Extreme Science and Engineering Discovery Environment (XSEDE), which is supported by National Science Foundation Grant No. OCI-1053575.  The author is supported by NASA grant NNX136AH77G.  The author thanks Alan Stern, Dave Hinson, Leslie Young, and Henry Throop for providing coordinates of the New Horizons temperature retrievals; Leslie Young for discussions regarding the subsurface module; Tim Michaels for discussion regarding the computational aspects of the model; Jake Simon for providing comments on the manuscript; and an anonymous referee for providing additional comments on the manuscript.







\appendix

\section{Surface and Subsurface Energy Model}\label{se:appendixA}
\textbf{The subsurface, assumed to be \htwoo (\citet{Cruikshank:1997} and concluded by \citet{stern:2015}, has a temperature specified following \citet{young:2012}:
\begin{equation}\label{eq:Tss}
\rho_{ss} c_{ss} \frac{\partial T_{ss}}{\partial t}=\kappa_{ss}\frac{\partial^2 T}{\partial z^2},
\end{equation}
where $\rho_{ss}$ is the density of the subsurface, $c_{ss}$ is the specific heat of the subsurface, $T_{ss}$ is the temperature of the subsurface, $z$ is now extended below the surface (to negative values), and $\kappa_{ss}$ is the thermal conductivity of the subsurface.  A subsurface with 25 layers is assumed.  The thermal inertia $\Gamma$ and the diurnal skin depth $Z$ are given by
\begin{equation}
\Gamma=\sqrt{\rho_{ss} c_{ss} \kappa_{ss}}
\end{equation}
and
\begin{equation}
Z=\frac{\kappa_{ss}}{\Gamma \sqrt{\Omega}},
\end{equation}
where $\Omega=1.13856\times 10^{-5}$~s$^{-1}$ is Pluto's rotation rate.  The depth of each layer is $\delta z=1/4 Z$, where the first layer begins at 1/2 $\delta z$.  The depth of the subsurface is thus 4.74~m.  Using these values, the subsurface dynamics captures daily oscillations but not yearly ones.  However, since these simulations are meant to be short-term the impact is assumed to be minimal on the final result.  Future long-term simulations will have a deeper subsurface.}

\textbf{Equation~(\ref{eq:Tss}) is a simple diffusion relationship.  The discrete form of equation~(\ref{eq:Tss}) is
\begin{equation}
\rho_{ss} c_{ss}\frac{T_j^{n+1}-T_j^n}{\Delta t}=\kappa_{ss}\frac{T^n_{j+1}-2T_j^n+T^n_{j-1}}{\Delta z^2},
\end{equation}
where $n$ indicates the time index and $j$ indicates the level.  Defining $\beta=\kappa_{ss}\Delta t/\rho_{ss}c_{ss}\Delta z^2$, the subsurface temperature is time marched by 
\begin{equation}
T_j^{n+1}=\beta T_{j+1}^n+\left(1-2\beta\right)T_j^n+\beta T_{j-1}^n.
\end{equation}
At the bottom level, it is assumed that no heat is being conducted from below, in which case the third term on the right hand side disappears and the ``2'' in the second term on the right hand side becomes ``1.''  Also, for the top level ($j=1$), $T_{j+1}^n$ is taken to be the surface temperature.  }

\textbf{The surface temperature, in discrete form, is given by
\begin{equation}\label{eq:Ts}
T_s^{n+1}=T_s^n+\frac{\Delta t}{c_{s} \Delta z \rho_{s}}\left[S(1-A_s)-\epsilon_s\sigma(T_s^n)^4-\frac{\kappa_s}{\Delta z}\left(T_s^n-T_1^n\right)\right]
\end{equation}
where $A_s$ is the surface albedo and $c_s$ and $\rho_s$ are the specific heat and density of the surface layer, which may either be seasonal N$_2$ frost or the bare subsurface.  Note that the subsurface does not sublimate or freeze in this model, only surface N$_2$.  The terms in equation~(\ref{eq:Ts}) from left to right are insolation $S$, emission by the surface with $\epsilon_s$ being the emissivity of the surface and $\sigma$ is the Stefan-Boltzmann constant, and conduction between the surface and the upper subsurface layer (where $\kappa_s$ is the conductivity of the surface).  The layer thickness $\Delta z$, is one fourth the skin depth $Z=\kappa_{ss}/(\Gamma_{ss}\sqrt{\Omega})$.  Here $\Gamma_{ss}$ is the thermal inertia of the subsurface, equal to $sqrt{\rho_{ss}c_{ss}\kappa_{ss}}$.}

\textbf{During the 30 Earth-year (0.12 Pluto-year) integration time, it was found that the surface and subsurface parameters did not greatly affect the results.  There is reason to be sceptical of this result; multi-Pluto year simulations are in progress.  For now, the surface and subsurface parameters are as follows: subsurface density and subsurface specific heat capacity (H$_2$O ice) are  936~kg~m$^{-3}$ and 960~J~kg$^{-1}$~K$^{-1}$, respectively, and for the seasonal N$_2(\beta)$ ice, the density, and specific heat 1000~kg~m$^{-3}$ and 1340~J~kg$^{-1}$~K$^{-1}$, respectively~\citep{spencer:1997}\footnote{A value of 960~J~kg$^{-1}$~K$^{-1}$ was erroneously used for the specific heat of H$_2$O ice.  The value reported by \citet{spencer:1997} is 350~J~kg$^{-1}$~K$^{-1}$ for ice at 40~K.  Given the insensitivity of other surface parameters to subsurface temperature discussed later, this value is not thought to significantly impact the results; however the \citet{spencer:1997} will be used in future work.}.  The conductivity of the subsurface is 4.88~J~m$^{-1}$~K$^{-1}$~s$^{-1}$.  Until such time that multi-Pluto year simulations can be performed and the surface frost distribution can be ``spun-up'' alongside a full atmospheric treatment, it is assumed, based on current understanding of Pluto at the time of these simulations, that thickness of the N$_2$ surface ice is 1~m thick.  This value was chosen such that there is sufficient N$_2$ surface ice such that it will not sublimate over the integration time but that the properties of the subsurface are still important terms in the surface heat equation.  The surface and subsurface temperature were initialized at the N$_2$ frost temperature and isothermal with depth.  The surface emissivity was assumed to be 1, the surface conductivity 0.2~J m$^{-1}$ K$^{-1}$ s$^{-1}$, and the surface albedo 0.2.  The parameter space explored by this survey was much larger, a set of cases with a surface albedo of 0.9 and another set of cases with a surface emissivity of 0.2.  All other parameters being equal, the results puzzlingly were practically independent of surface albedo and emissivity.}

\textbf{In the GCM results, the surface temperature remained at the N$_2$ freezing temperature, globally for all cases.  Over the course of every simulation, N$_2$ ice was found to accumulate everywhere on the surface.  No gridpoints sublimated the initial surface N$_2$ ice at any time during the simulations.  We expect a condensation flow (i.e. transport of volatiles by wind), but such a flow was not identified. It is likely that the surface needs far more than 30 Earth years to equilibrate, as shown by other modelling studies~\citep[e.g.][]{hansen:1996,young:2013}.  The thickness of ice that accumulated was of order 1~mm.  No structure was found in the ice accumulation vs. latitude and longitude.  The subsurface temperature changed by less than 0.01~K at any depth globally.  Simulations in progress with integration times of several Pluto years will be the subject of future work.  Moreover, the subsurface and surface model and the behaviour of the volatile scheme will be tested with a simplified atmosphere to confirm agreement with surface energy balance models \citep[e.g.][]{hansen:1996,young:2013}. }

\section{Additional Figures and Discussion for Cases 1, 2, and 4}\label{se:appendixB}

\begin{figure*}

 \noindent\includegraphics[width=1.0\textwidth]{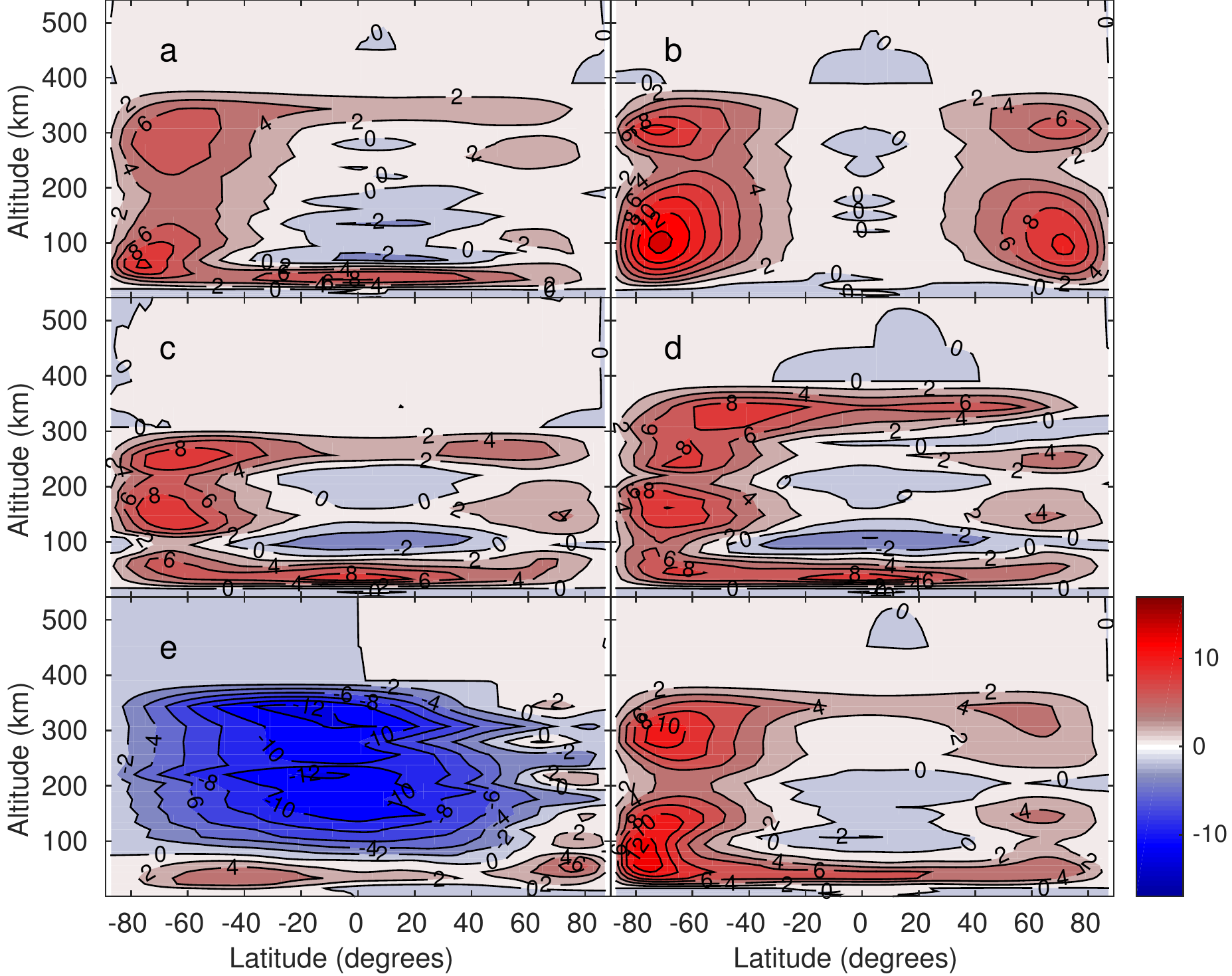}

 \caption{\label{fg:extraplots335}  Same as Fig.~\ref{fg:extraplots369} but for Case 1.   The panel with the adjacent colorbar is the same as Fig.~\ref{fg:u1}, top left panel for comparison.}
 
  \end{figure*}
\begin{figure*}
\noindent\includegraphics[width=1.0\textwidth]{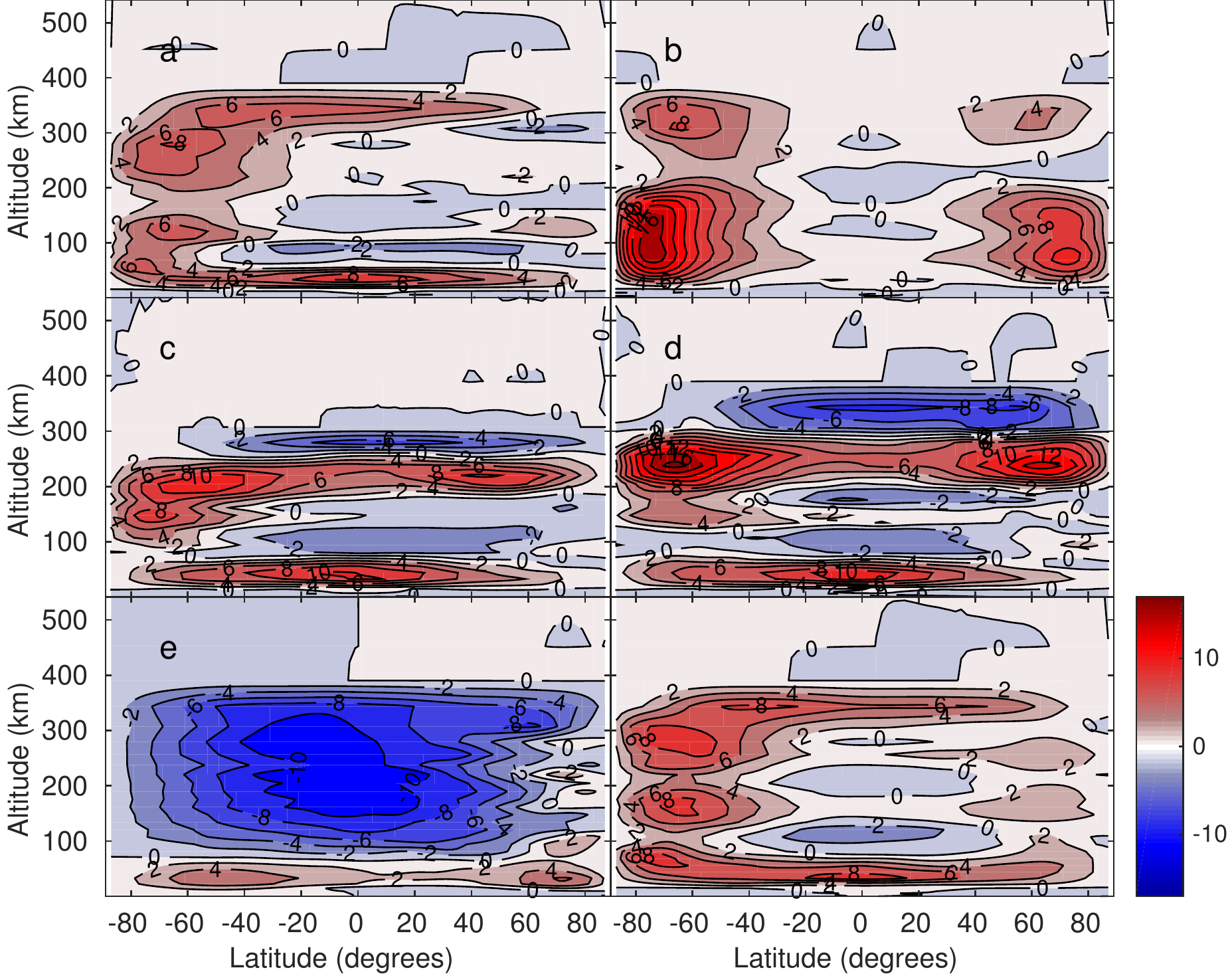}

 \caption{\label{fg:extraplots343}  Same as Fig.~\ref{fg:extraplots369} but for Case 2.   The panel with the adjacent colorbar is the same as Fig.~\ref{fg:u1}, top right panel for comparison.}

  \end{figure*}
\begin{figure*}

 \noindent\includegraphics[width=1.0\textwidth]{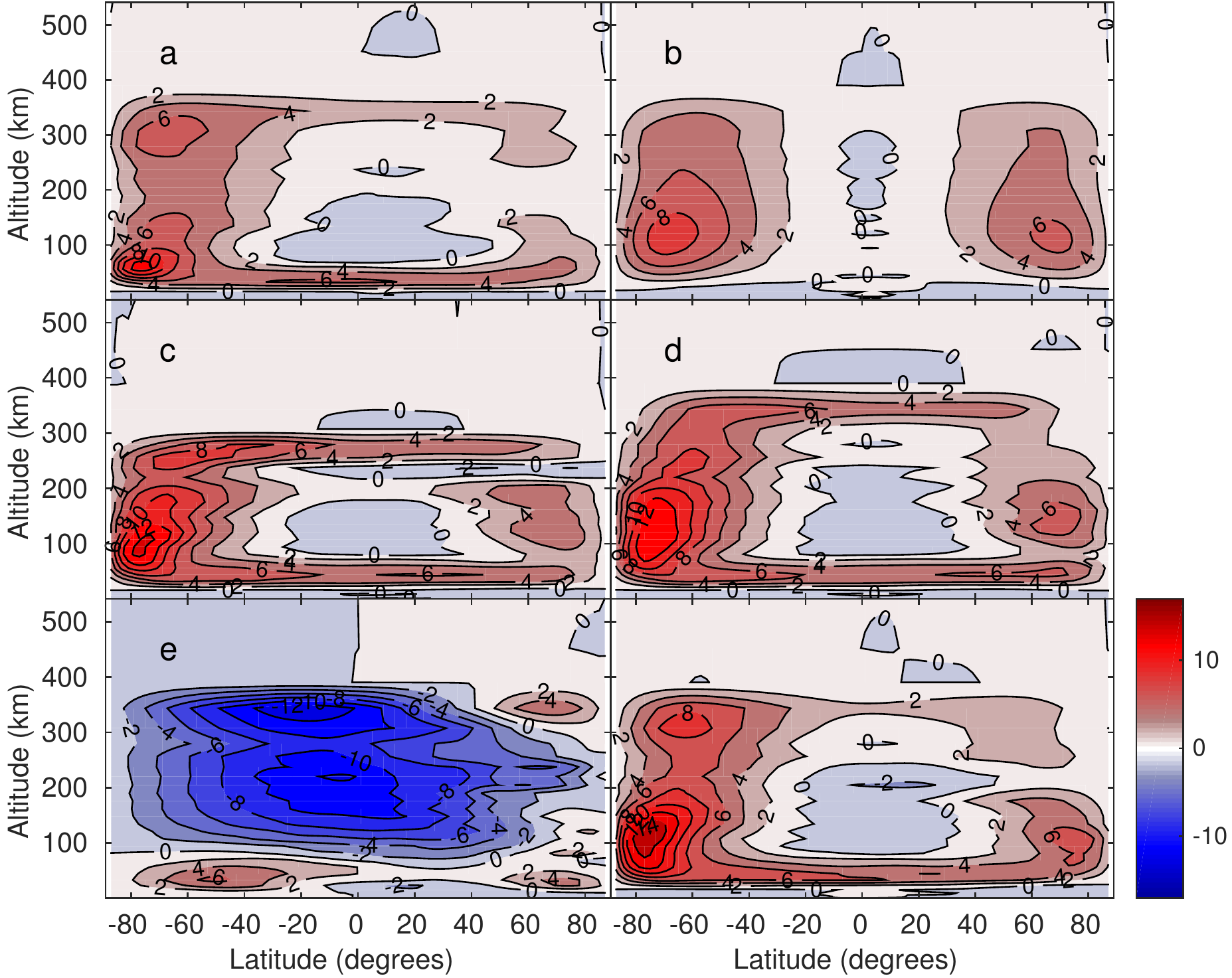}

 \caption{\label{fg:extraplots370} Same as Fig.~\ref{fg:extraplots369} but for Case 4.   The panel with the adjacent colorbar is the same as Fig.~\ref{fg:u1}, bottom right panel for comparison.}

  \end{figure*}
\begin{figure*}
 \noindent\includegraphics[width=0.75\textwidth]{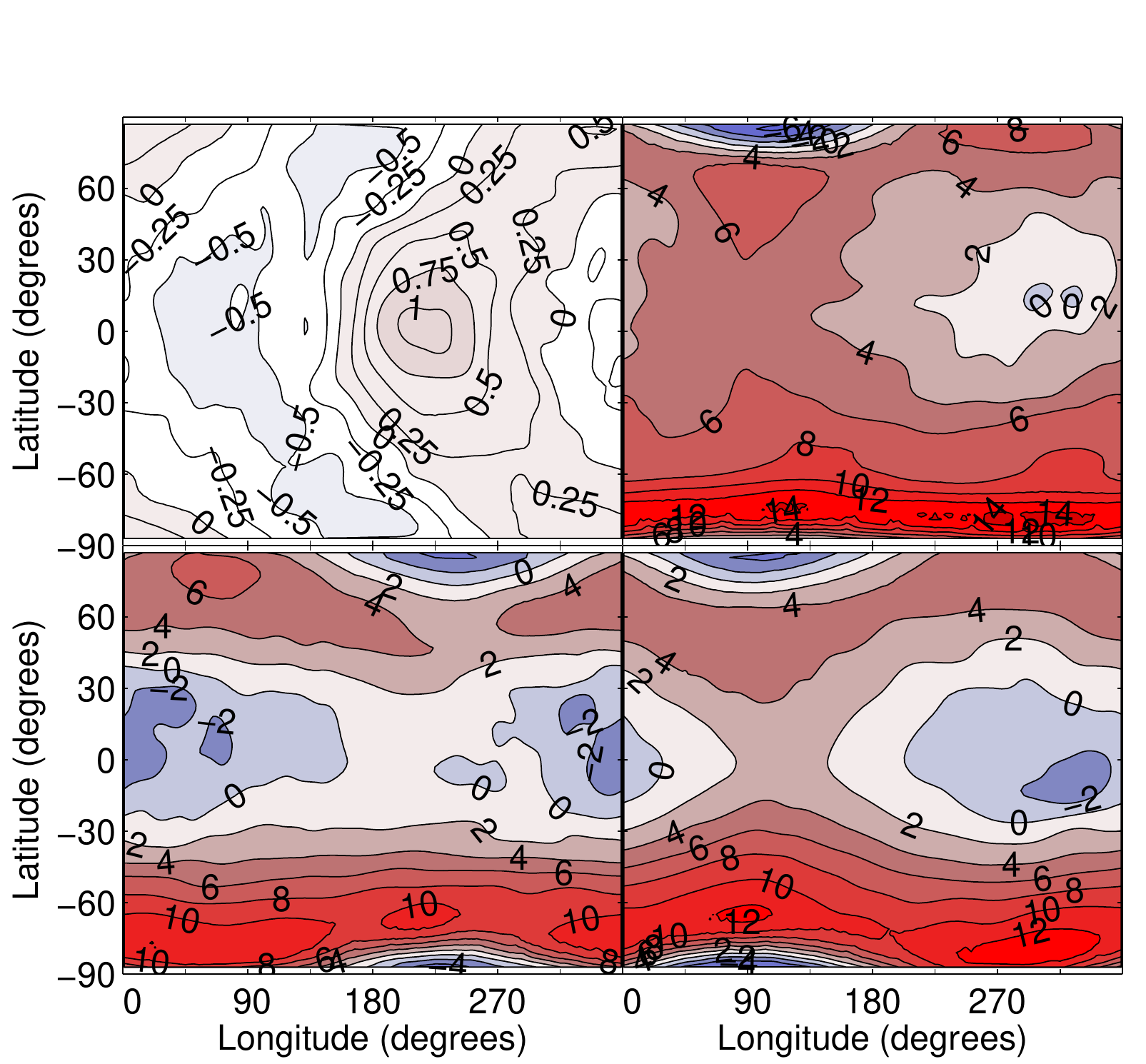}

\caption{\label{fg:uxy1}  Same as Fig.~\ref{fg:uxy3}, but for Case 1.  The altitudes of the panels from left to right and top to bottom are 2.4, 37, 109, and 248~km.  (The difference from the other simulations results from the conversion from pressure to altitude).}

  \end{figure*}
\begin{figure*}
 \noindent\includegraphics[width=0.75\textwidth]{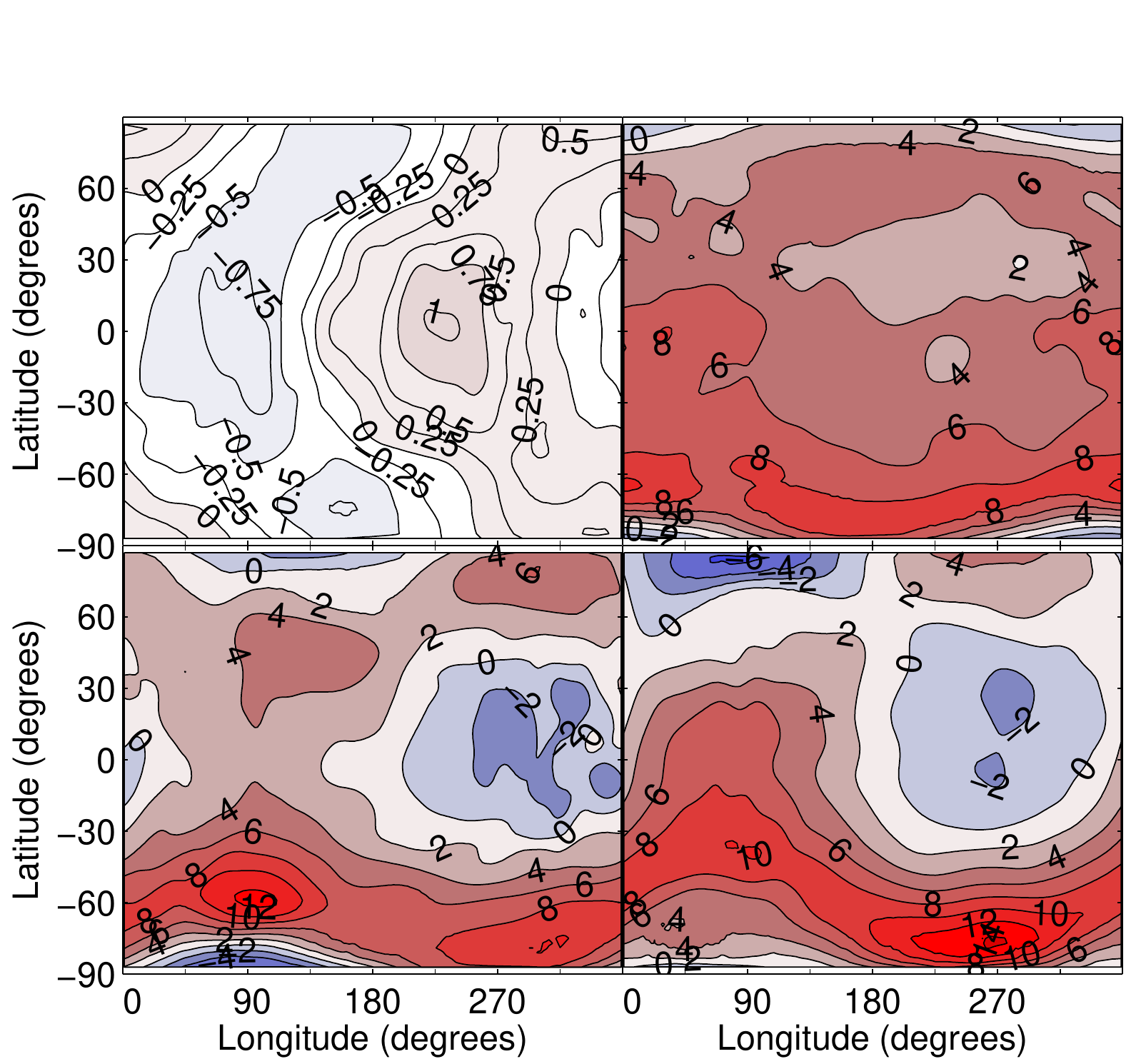}

 \caption{\label{fg:uxy2}  Same as Fig.~\ref{fg:uxy3}, but for Case 2.  The altitudes of the panels from left to right and top to bottom are 2.5, 38, 112, and 249~km.  (The difference from the other simulations results from the conversion from pressure to altitude).}

  \end{figure*}
\begin{figure*}

 \noindent\includegraphics[width=0.75\textwidth]{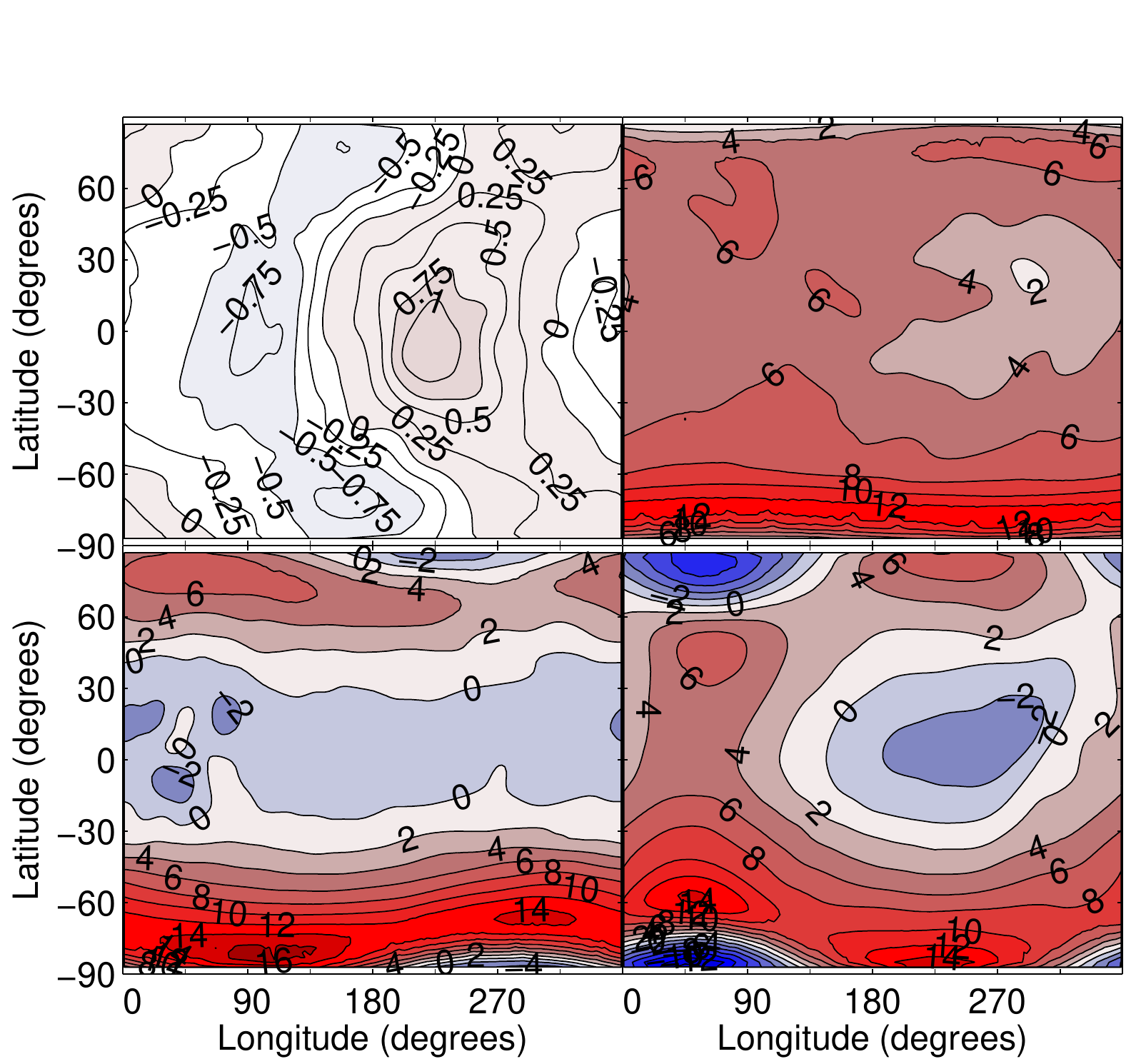}

 \caption{\label{fg:uxy4}  Same as Fig.~\ref{fg:uxy3}, but for Case 4. The altitudes of the panels from left to right and top to bottom are 2.4, 35, 105, and 231~km.  (The difference from the other simulations results from the conversion from pressure to altitude).}
 
  \end{figure*}
\begin{figure*}
 \noindent\includegraphics[width=0.75\textwidth]{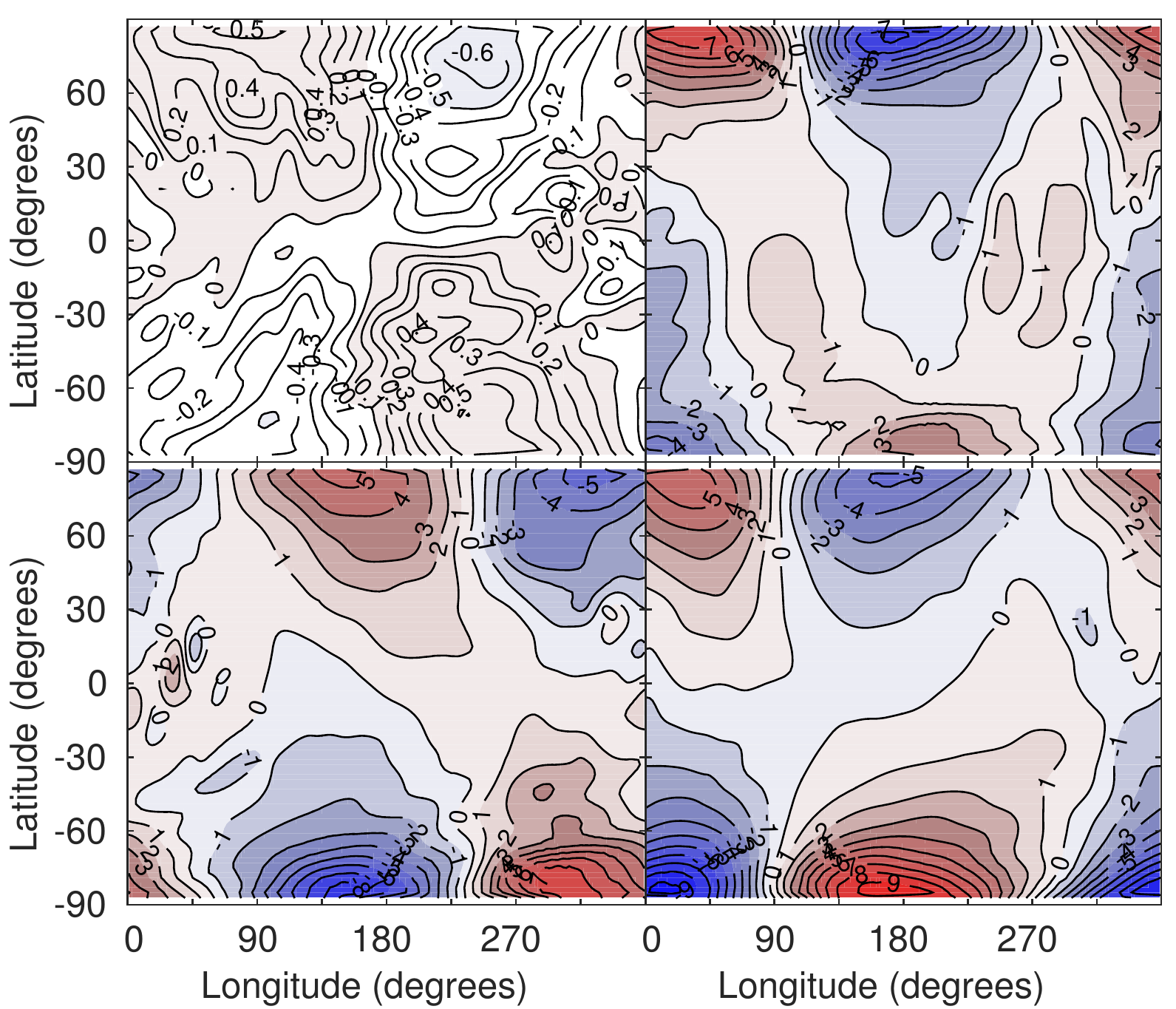}

\caption{\label{fg:vxy1}   Same as Fig.~\ref{fg:vxy3}, but for Case 1.  The altitudes of the panels from left to right and top to bottom are 2.4, 37, 109, and 248~km.  (The difference from the other simulations results from the conversion from pressure to altitude).}

  \end{figure*}
\begin{figure*}
 \noindent\includegraphics[width=0.75\textwidth]{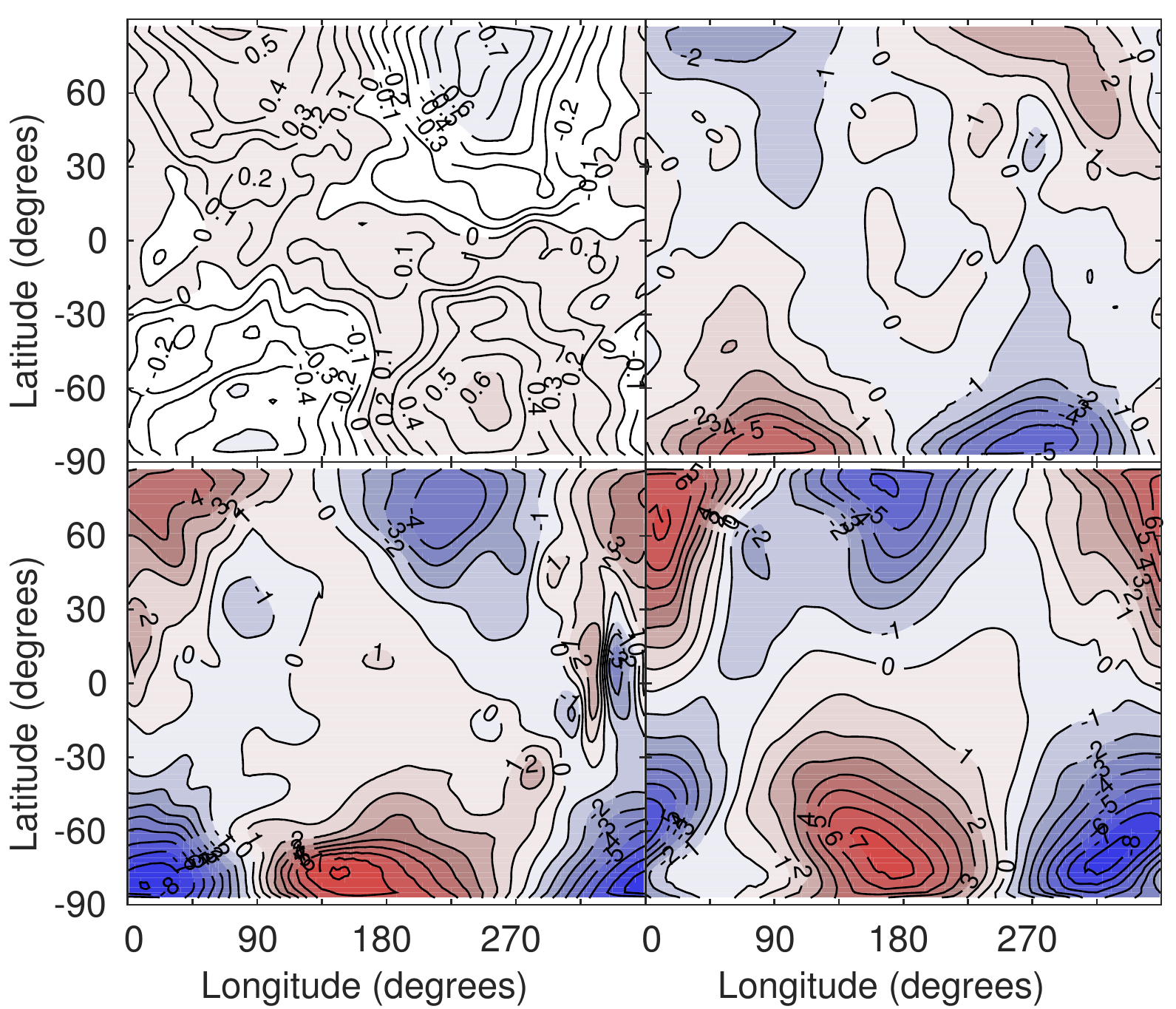}

 \caption{\label{fg:vxy2}  Same as Fig.~\ref{fg:vxy3}, but for Case 2.  The altitudes of the panels from left to right and top to bottom are 2.5, 38, 112, and 249~km.  (The difference from the other simulations results from the conversion from pressure to altitude).}

  \end{figure*}
\begin{figure*}

 \noindent\includegraphics[width=0.75\textwidth]{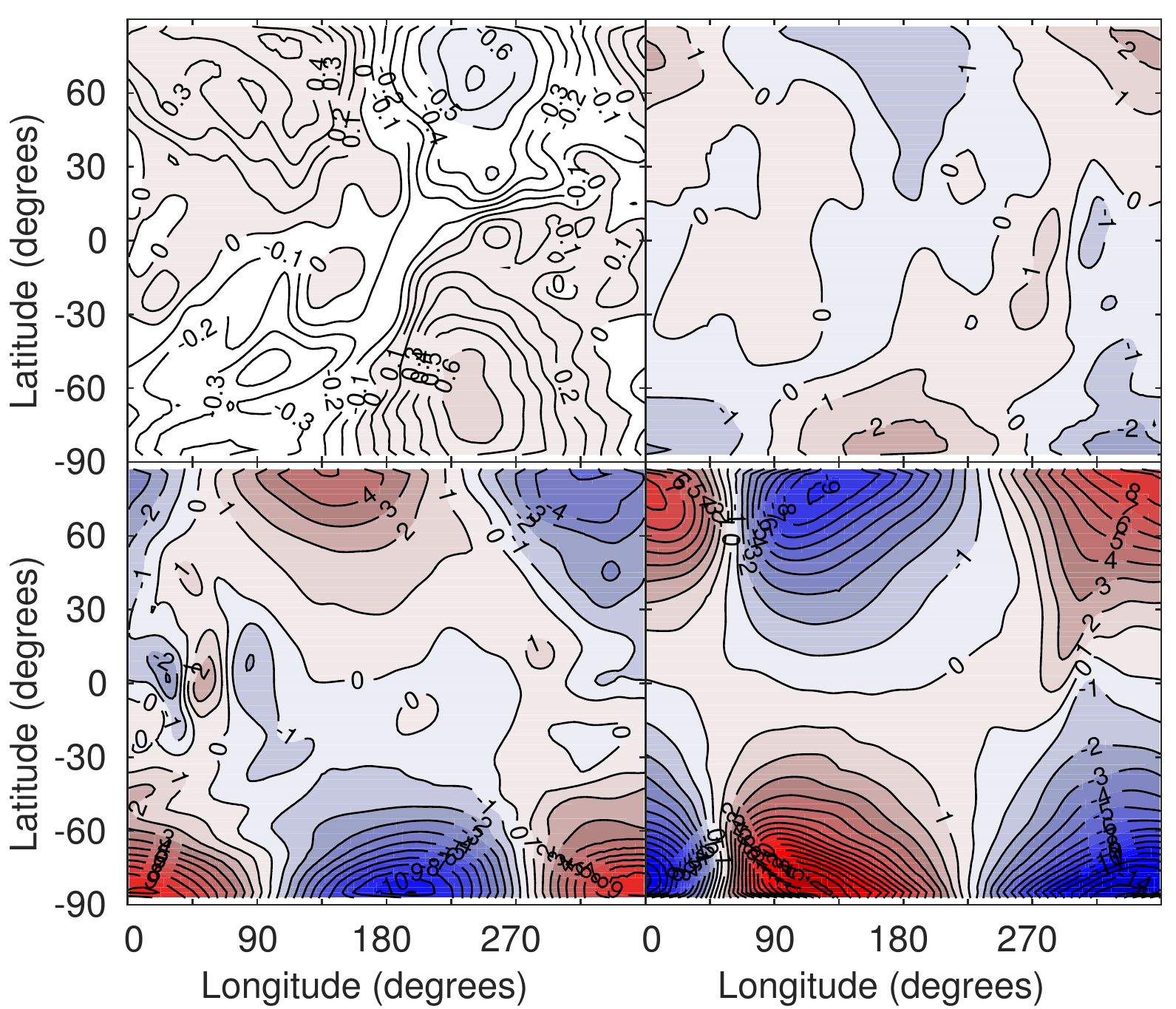}

 \caption{\label{fg:vxy4}  Same as Fig.~\ref{fg:vxy3}, but for Case 4. The altitudes of the panels from left to right and top to bottom are 2.4, 35, 105, and 231~km.  (The difference from the other simulations results from the conversion from pressure to altitude).}
  \end{figure*}
\begin{figure*}
 \noindent\includegraphics[width=0.75\textwidth]{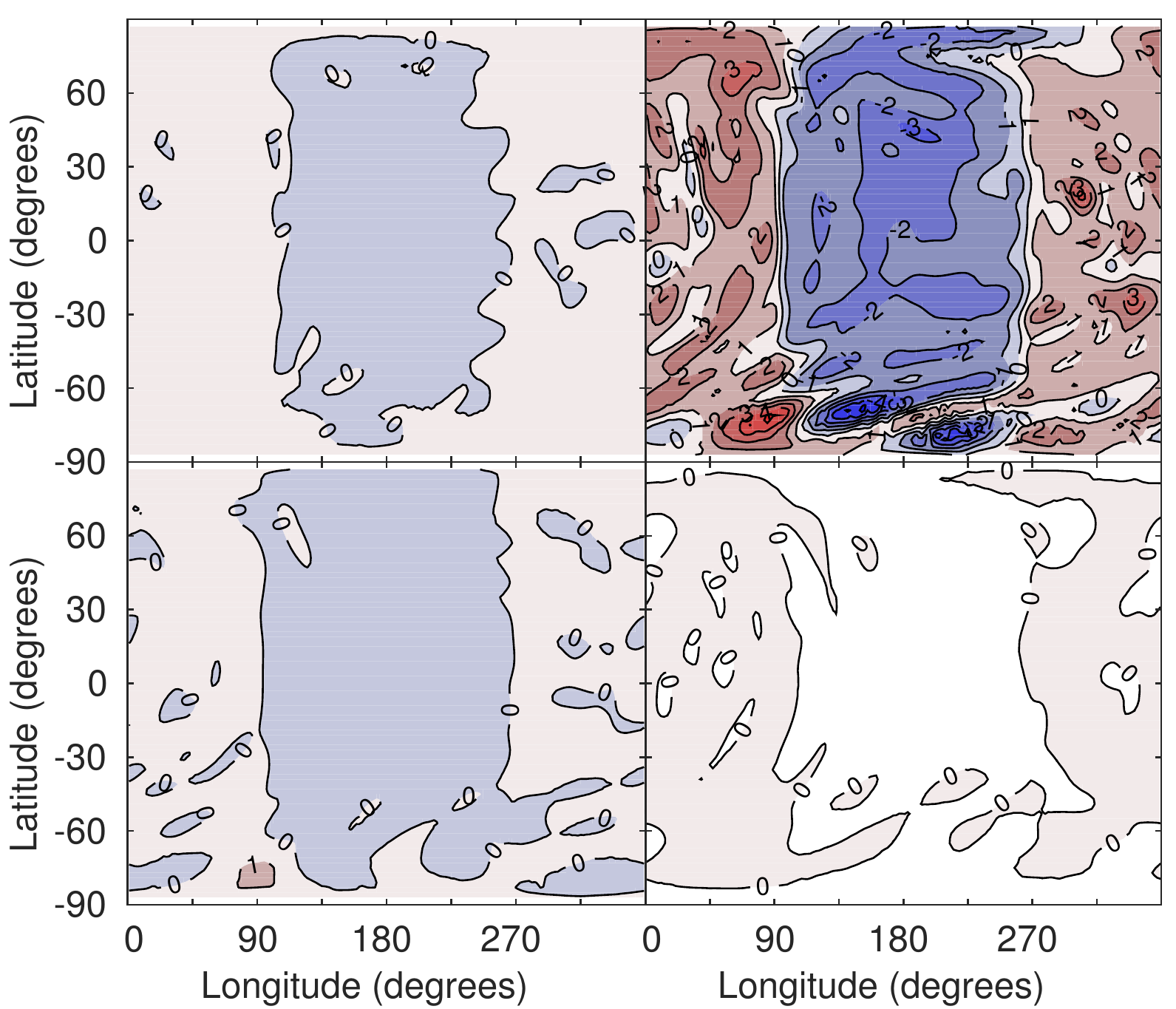}

\caption{\label{fg:wxy1}  Same as Fig.~\ref{fg:wxy3}, but for Case 1.  The altitudes of the panels from left to right and top to bottom  are 2.4, 37, 109, and 248~km.  (The difference from the other simulations results from the conversion from pressure to altitude).}

  \end{figure*}
\begin{figure*}
 \noindent\includegraphics[width=0.75\textwidth]{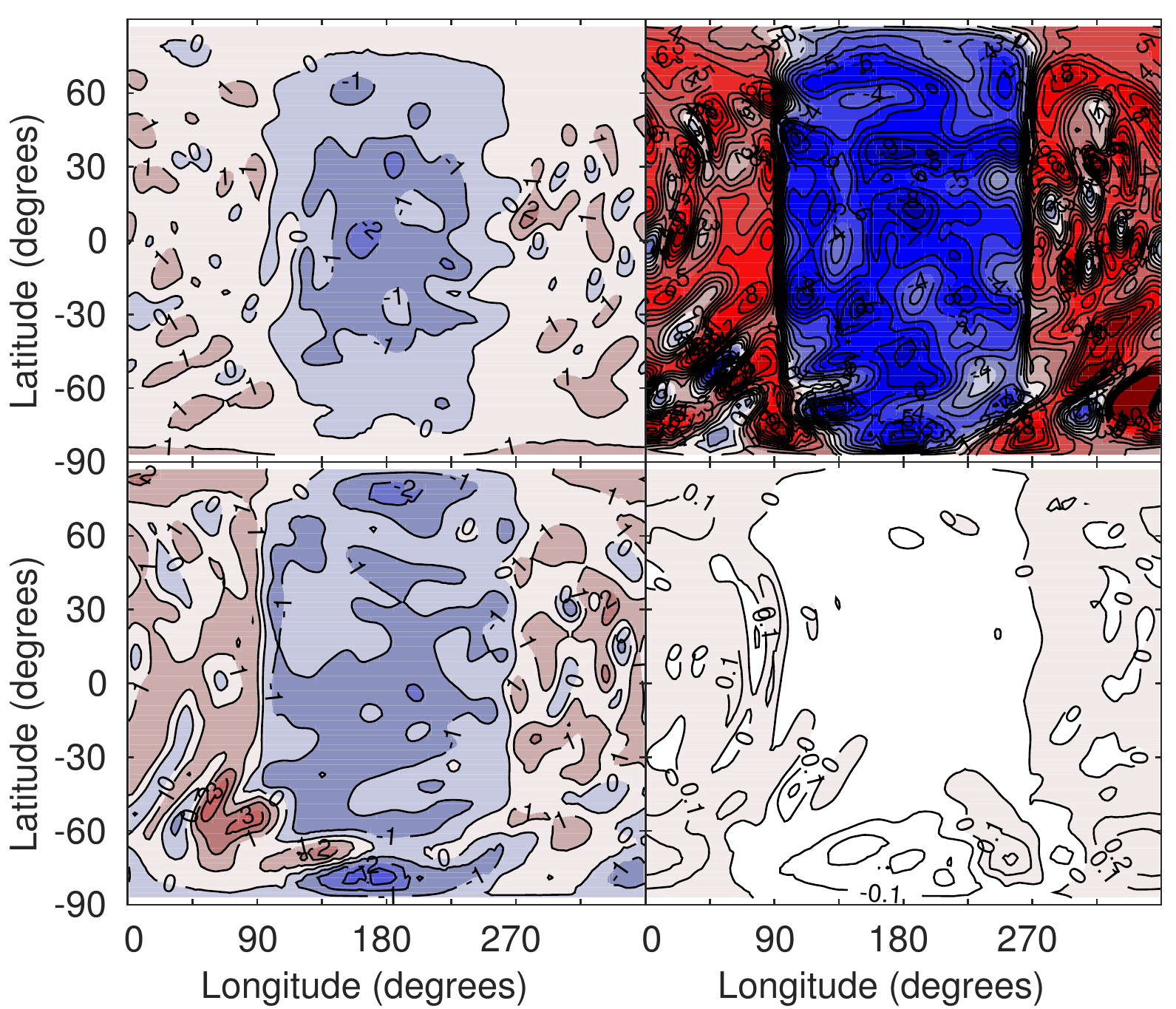}

 \caption{\label{fg:wxy2}  Same as Fig.~\ref{fg:wxy3}, but for Case 2.  The altitudes of the panels from left to right and top to bottom  are 2.5, 38, 112, and 249~km.  (The difference from the other simulations results from the conversion from pressure to altitude).}

  \end{figure*}
\begin{figure*}

 \noindent\includegraphics[width=0.75\textwidth]{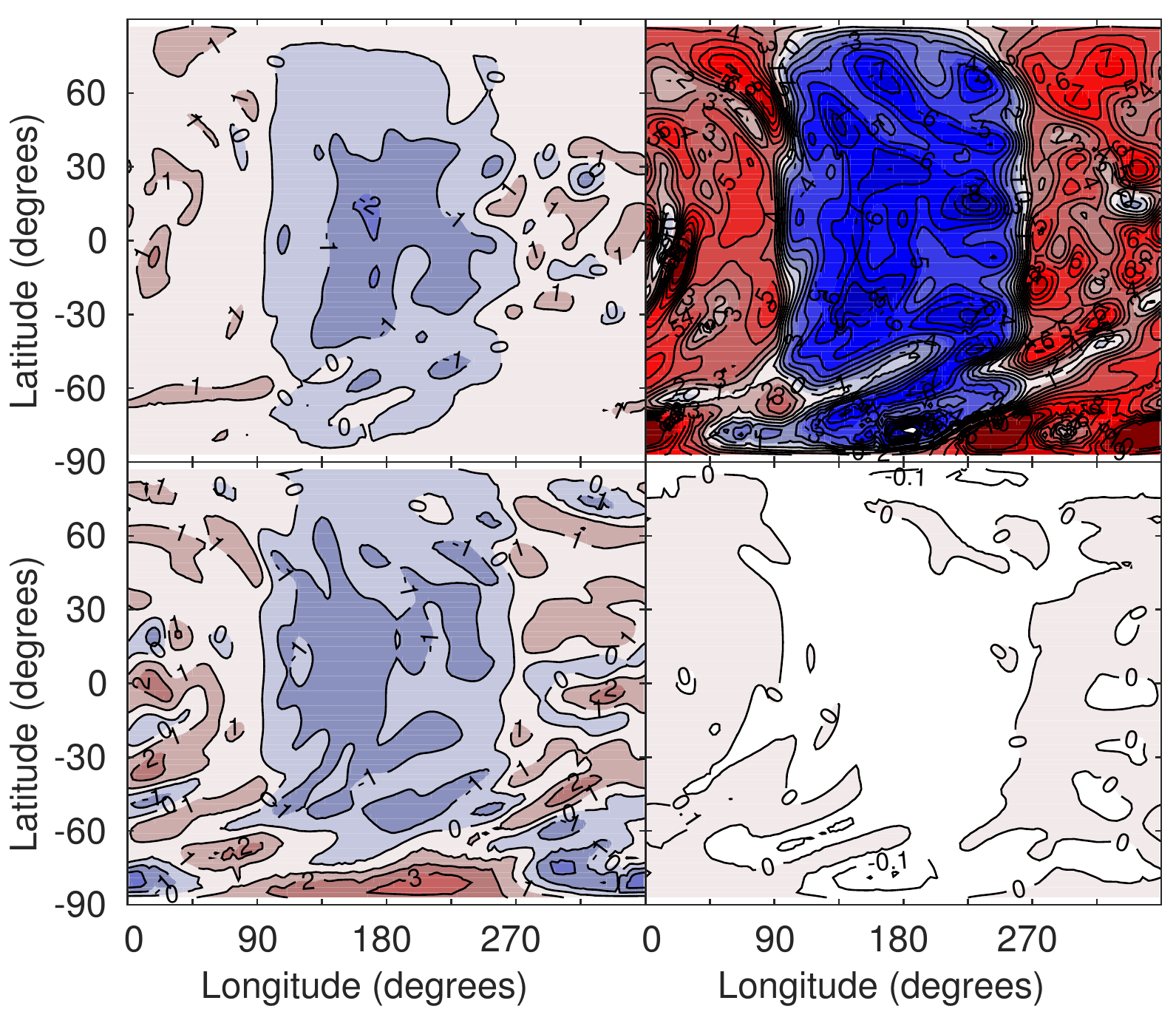}

 \caption{\label{fg:wxy4}  Same as Fig.~\ref{fg:wxy1}, but for Case 4. The altitudes of the panels from left to right and top to bottom are 2.4, 35, 105, and 231~km.  (The difference from the other simulations results from the conversion from pressure to altitude).}
 
  \end{figure*}

Simulations with the the global mean surface pressures and \methane~mixing ratios for Cases 1, 2, and 4 are worth examining in the came close detail as Case 3 in the main part of the results and discussion, because they potentially represent states that Pluto's atmosphere may have existed in even as recently as the current observational epoch (from 1988 onwards).  Figures~\ref{fg:extraplots335}--\ref{fg:extraplots370} show results from the simulations in which one feature of the model (CO cooling, 2.3~$\mu$m \methane~band, upper atmosphere drag coefficient, and volatile cycle) was changed, as in Fig.~\ref{fg:extraplots369}.  Figures~\ref{fg:uxy1}--\ref{fg:uxy4} show latitude-longitude plots of the longitudinal wind at different atmospheric heights, as in Fig.~\ref{fg:uxy3}.  Figures~\ref{fg:vxy1}--\ref{fg:vxy4} and Figs.~\ref{fg:wxy1}--\ref{fg:wxy4} show latitude-longitude plots of the latitudinal and vertical wind, respectively, at different atmospheric heights, as in Figs.~\ref{fg:vxy3} and \ref{fg:wxy3}, respectively. 

\bsp	
\label{lastpage}
\end{document}